\documentclass[twocolumn,showpacs,preprintnumbers,amsmath,amssymb,prx,superscriptaddress,longbibliography,amsfonts]{revtex4-1} 
\pdfoutput=1

\usepackage[pdftex]{graphicx} 

\usepackage{color}

\usepackage{epsfig,amsopn}
\usepackage{graphicx}
\usepackage{amsthm}
\usepackage{enumerate}
\usepackage{ctable}
\usepackage{epstopdf}
\usepackage{amsmath,amssymb,natbib,bm,graphicx,url}
\usepackage{psfrag}
\usepackage{amssymb}
\usepackage{subfigure}
\usepackage{color}
\usepackage{wasysym}
\usepackage{float}
\usepackage{soul}

\newcommand{\be}{\begin{equation}}
\newcommand{\ee}{\end{equation}}
\newcommand{\bea}{\begin{eqnarray}}
\newcommand{\eea}{\end{eqnarray}}

\newcommand{\im}{\text{Im}}
\newcommand{\re}{\text{Re}}

\begin{document}

\title{Hall Effect Gyrators and Circulators}

\author{Giovanni Viola}
\affiliation{Institute for Quantum Information, RWTH Aachen University, D-52056 Aachen, Germany}
\author{David P. DiVincenzo}
\affiliation{Institute for Quantum Information, RWTH Aachen University, D-52056 Aachen, Germany}
\affiliation{Peter Gr{\"u}nberg Institute: Theoretical Nanoelectronics, Research Center J{\"u}lich}
\affiliation{J{\"u}lich-Aachen Research Alliance (JARA), Fundamentals of Future Information Technologies}

\begin{abstract}
The electronic circulator, and its close relative the gyrator, are invaluable tools for noise management and signal routing in the current generation of low-temperature microwave systems for the implementation of new quantum technologies.  The current implementation of these devices using the Faraday effect is satisfactory, but requires a bulky structure whose physical dimension is close to the microwave wavelength employed.  The Hall effect is an alternative non-reciprocal effect that can also be used to produce desired device functionality.  We review earlier efforts to use an ohmically-contacted four-terminal Hall bar, explaining why this approach leads to unacceptably high device loss.  We find that capacitive coupling to such a Hall conductor has much greater promise for achieving good circulator and gyrator functionality.  We formulate a classical Ohm-Hall analysis for calculating the properties of such a device, and show how this classical theory simplifies remarkably in the limiting case of the Hall angle approaching 90 degrees.  In this limit we find that either a four-terminal or a three-terminal capacitive device can give excellent circulator behavior, with device dimensions far smaller than the a.c. wavelength.  An experiment is proposed to achieve GHz-band gyration in millimetre (and smaller) scale structures employing either semiconductor heterostructure or graphene Hall conductors.  An inductively coupled scheme for realising a Hall gyrator is also analysed.     
\end{abstract}

\date{\today}

\pacs{71.10.Pm}
\maketitle

\section{Introduction}
\begin{figure}[b]
\centering
\includegraphics[scale=0.3]{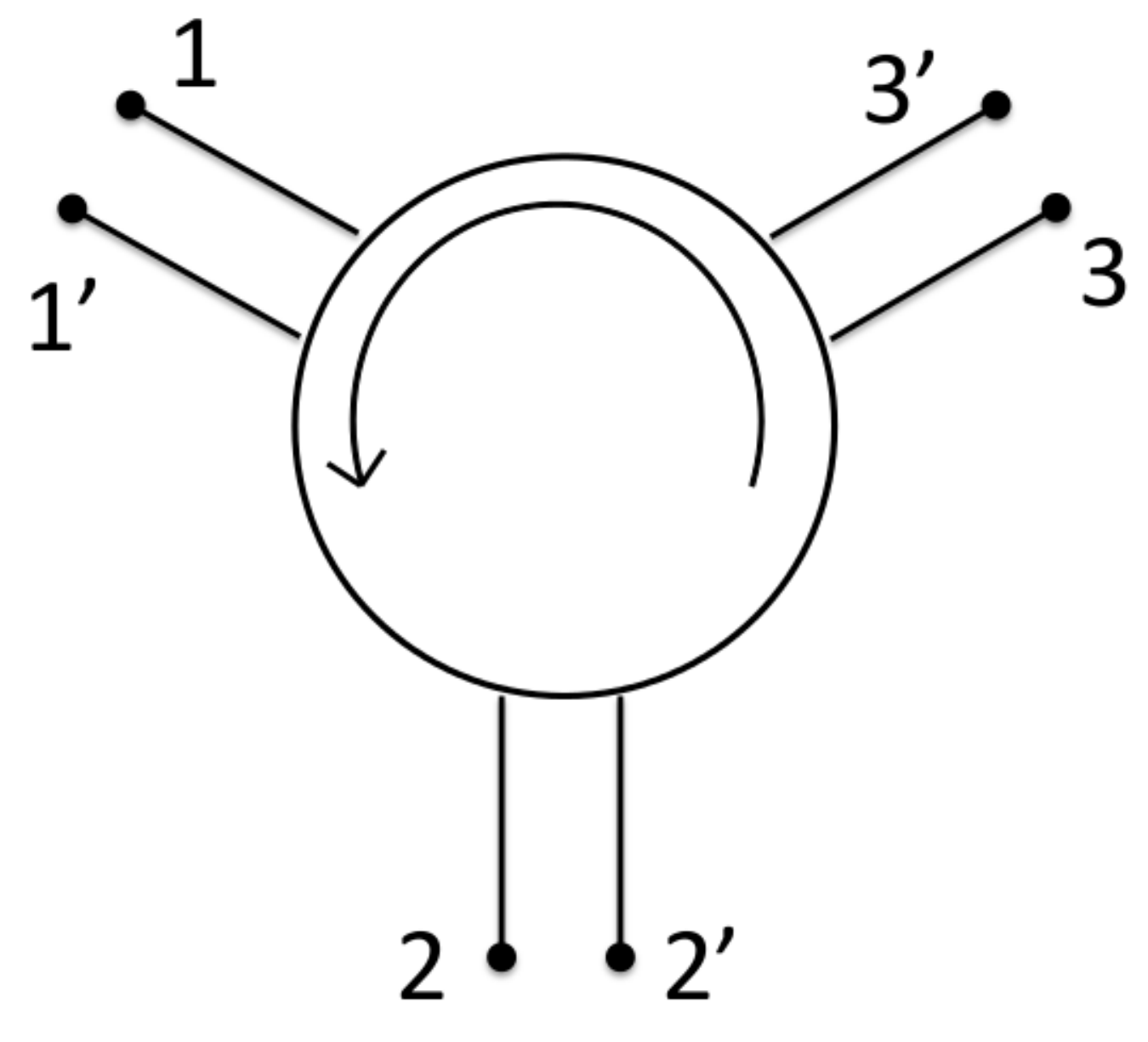}
\caption{The conventional symbol for the three-port circulator~\cite{Pozar}, indicating counterclockwise circulation ($1\rightarrow 2\rightarrow 3$).  We adopt an older variant of the notation~\cite{Newcomb} in which the ports are depicted as terminal pairs, emphasising that the input signal to a port can be characterised by the voltage difference between two discrete nodes.  The terminal pair of a port will always have a primed and unprimed label.  While the primed terminal can often be indicated as ``ground'', it is not necessarily the case that there is a d.c. connection between the primed terminals inside the device.  The port condition on the currents is that the current into the unprimed terminal is equal to the current out of the corresponding primed terminal.}
\label{fig:f1}
\end{figure}

The Faraday-effect circulator is an unsung workhorse of the contemporary surge of low temperature microwave device physics, playing a key role in permitting low noise control and measurement of superconducting qubits and resonators.  The essence of the three-port circulator is its non-reciprocal routing of signals: electromagnetic radiation is passed cyclically from one port to its neighbor -- radiation in at port one goes out at port two, in at two goes out at three, and in at three goes out at one, see Fig.~\ref{fig:f1}. The $S$ matrix describing the circulator is simply~\cite{Pozar}
\begin{equation}
S=\left(\begin{array}{lll}0&0&1\\1&0&0\\0&1&0\end{array}\right)\label{theS}.
\end{equation}  

\begin{figure}[htp]
\centering
\includegraphics[scale=0.17]{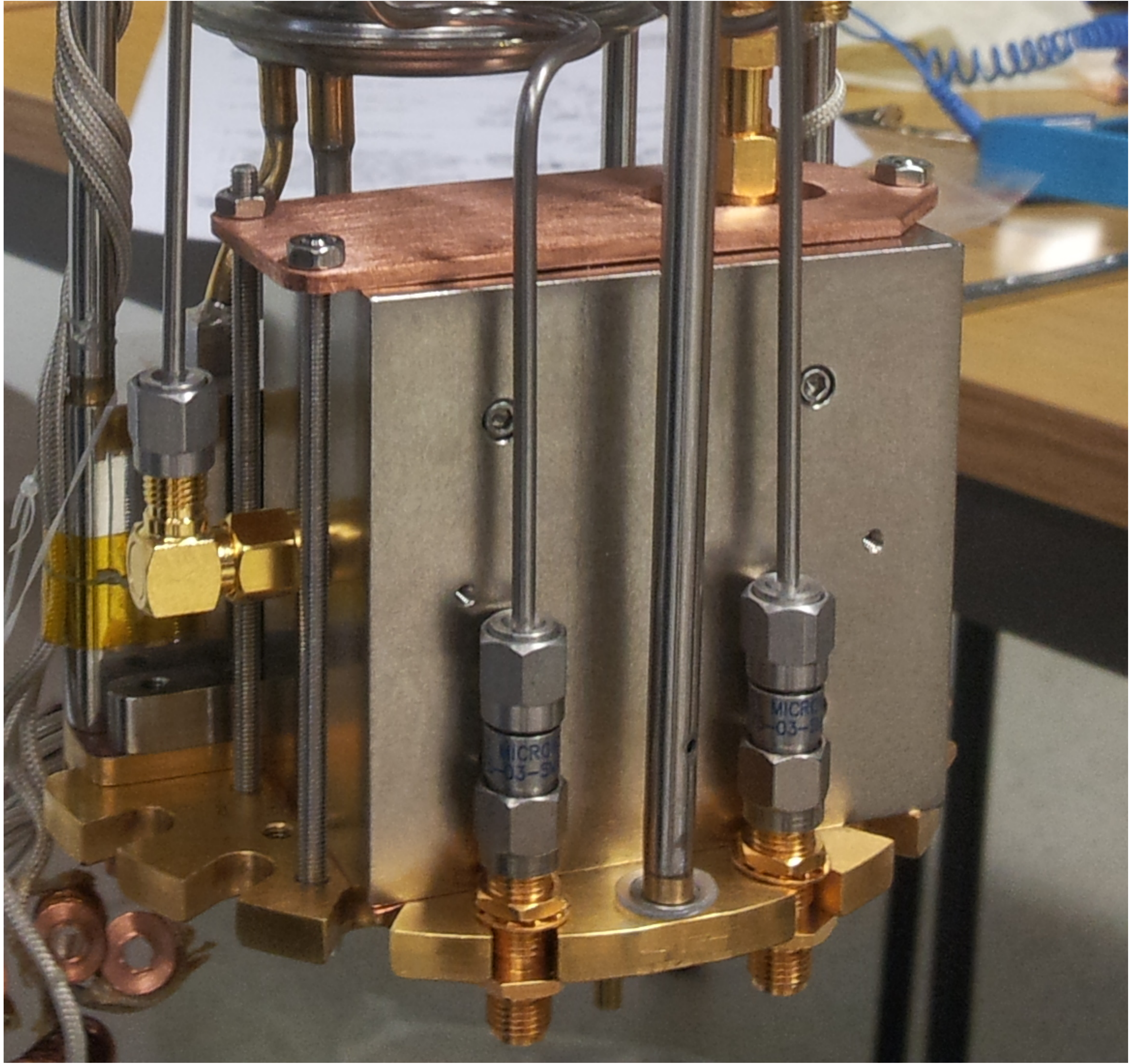}
\caption{A standard Faraday circulator mounted in a low-temperature experiment.  The circulator, with its enclosing magnetic shield, is contained in the rectangular steel-colorerd box (dimensions 6cm$\times$6cm$\times$3cm).  This circulator is designed for an operating frequency of 400MHz, those in use for GHz-band experiments are somewhat more compact.  Another interesting view of microwave circulators in action in a multi-qubit experiment can her found in~\cite{1402.1868P}. In the present paper we propose a scheme for achieving a passive circulator with a physical scale 10-100 times smaller.  As in the present day circulator, significant magnetic fields (perhaps approaching Tesla scale) will be required.  This will require a shield structure as in the device shown, but again proportionally smaller.}
\label{fig:f2x}
\end{figure}

Here and later, the $S$ matrix relates the incoming amplitudes of electromagnetic waves to the outgoing amplitudes.  We are {\em not} referring to the quantum $S$ matrix for electronic wave function amplitudes.

Fig.~\ref{fig:f2x} shows a circulator in place in a contemporary qubit experiment~\footnote{From the lab of H. Bluhm, RWTH Aachen.}.
A typical present-day experiment involving just a single superconducting qubit requires no fewer than four circulators~\cite{Riste12} for the proper management of signals used to do high-fidelity, rapid measurements on the qubit.  (During the writing of this manuscript, a three-qubit experiment was reported~\cite{12circulator} with no fewer than eleven circulators!) While highly reliable and reasonably close to ideal in their designed frequency band of operation, they are quite bulky.  The few-centimeter linear dimension of a circulator operating in the few-gigahertz frequency range is explained very simply: the Faraday effect causes circulation by a wave-interference phenomenon~\cite{Pozar}, requiring a physical scale on the order of the wavelength.  Naive scaling of experiments to, say, hundreds of qubits would require an impractically large volume of low-temperature space devoted to circulators.  A primary objective of the present work has been to identify a new physical basis for the circulator which permits very significant miniaturization.

It is well known how to achieve miniaturatization of the circulator function to far below wavelength scale using operational amplifier circuits~\cite{Berndt69}.  But even if such electronic circuits could work at cryogenic temperatures, their power dissipation and noise performance would be unacceptable for current applications.  Very compact gyrators may be achievable with SQUID structures, according to a preliminary study~\cite{PhysRevA.82.043811}.  Another form of active gyrator that is workable at low temperature is currently considered in the area of parametric Josephson devices~\cite{Irwin}; while these are integrated-circuit devices~\cite{PhysRevX.3.031001}, their employment of resonator structures~\cite{abdoarxiv,PhysRevLett.110.173902,Bergeal2010,Bergeal2010a} means that their compactness is not guaranteed.  These schemes seem to be related to theoretical ideas from long ago for realising non-reciprocal devices with parametrically modulated linear components~\cite{Anderson1,Anderson2,Anderson3}; these theories were apparently never put into practice (see also p. 30 of~\cite{Newcomb}).    

\begin{figure}[htp]
\centering
\includegraphics[scale=0.37]{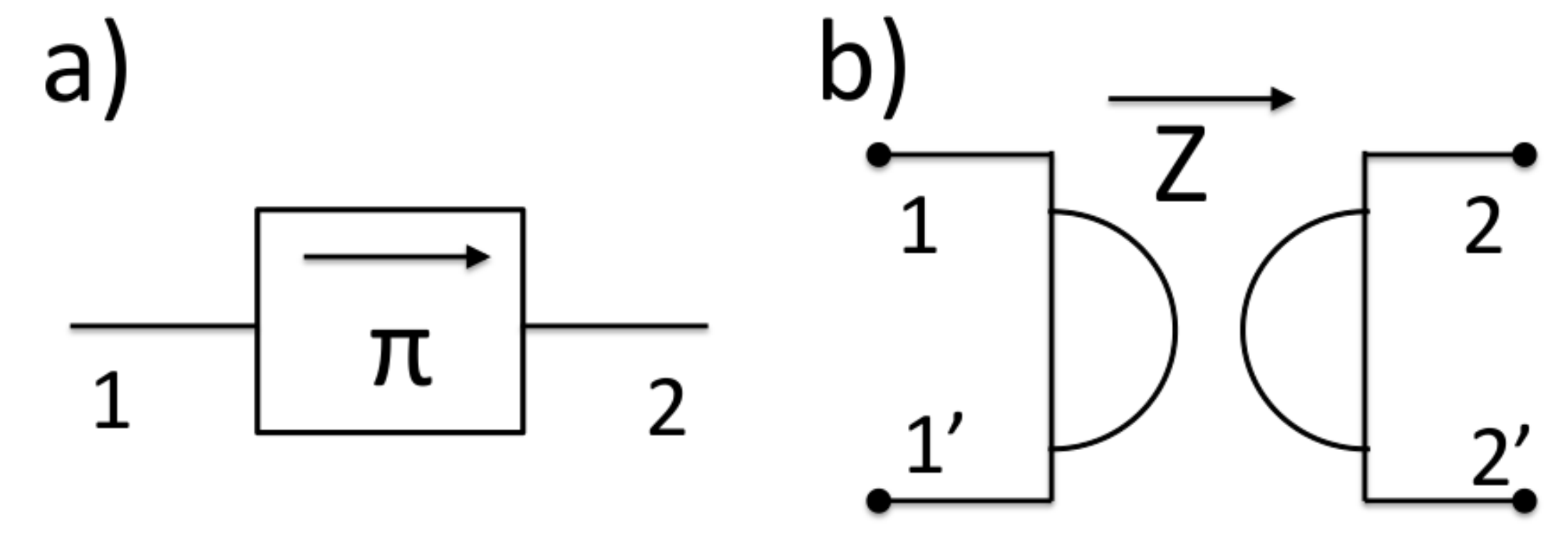}
\caption{The gyrator, a two-port non-reciprocal device.  a) The ``quasi-optical'' conventional symbol~\cite{Pozar}, emphasising that that the gyrator imparts a phase inversion to signals propagating in one direction only. b) The lumped-element 4-terminal symbol of the gyrator.  According to Tellegen's definition~\cite{TellegenPRR}, the port currents and voltages satisfy the equations $V_2=ZI_1,~ V_1=-ZI_2$. $Z$ is the gyration resistance.}
\label{fig:f2}
\end{figure}

Several devices are straightforward derivatives of the circulator, which deserve consideration in the own right.  If one of the ports of the circulator is terminated in a matched load, that is, with a resistance $Z_0$ equal to the wave or source impedance chosen for signal propagation in the system (often 50 or 100$\Omega$, this is just the ratio of the guided wave's voltage to its current), one obtains the {\em isolator}, in which signals are perfectly transmitted in one direction between the remaining ports, and perfectly absorbed in the other direction.  In a recent experiment~\cite{Riste12} three of the four circulators are configured as isolators, serving the function of blocking noisy (high temperature) radiation from entering the low-temperature part of the experiment.  If one port is unterminated, the resulting device is uninteresting: the reverse-direction signal undergoes reflection at the open port and is transmitted identically to transmission in the other direction -- both forward- and reverse-propagating signals are transmitted without modification.  If one circulator port is terminated in a short circuit, the resulting highly non-trivial two-port device is known as a {\em gyrator}.  It is a maximally non-reciprocal device in its effect on the phase of signals.  Forward-directed signals acquire no phase shift, while back-propagating signals are inverted, being phase shifted by $\pi$, coming from the phase-inverting reflection at the short-circuit termination.  This is indicated in the conventional symbol for the gyrator (Fig.~\ref{fig:f2} (a)).  

The gyrator is arguably more fundamental than the circulator, and it will be the main focus of study in this paper.  Historically the gyrator predates the circulator and was responsible for its discovery.  There are two distinct methods, to be reviewed below, for obtaining circulator action using a gyrator.  Before introducing these, it is best that we first summarize a few of the basic points about the mathematical description of the gyrator~\cite{Pozar,Newcomb}.  The scattering matrix of the ideal gyrator is~\cite{Pozar}
\begin{equation}
S=\left(\begin{array}{rr}0&-1\\1&0 \end{array}\right).\label{gyS}
\end{equation}
The impedance and admittance matrices of the ``matched'' gyrator, for which the internal impedance and the source impedance both equal $Z_0$, are given by the standard matrix formulas~\cite{Pozar}
\begin{equation}
Z=Z_0(I+S)(I-S)^{-1}=Z_0\left(\begin{array}{rr}0&-1\\1&0 \end{array}\right).\label{gyZ}
\end{equation}
\begin{equation}
Y=Z^{-1}=-{1\over Z_0}\left(\begin{array}{rr}0&-1\\1&0 \end{array}\right)\label{gyY}.
\end{equation}
Port impedances or admittances become especially useful quantities in the ``near field'' or ``lumped device'' limit, when the device dimensions are much smaller than the wavelength of interest~\cite{Pozar}.  This is not particularly true of the present-day Faraday devices, but will be true for the devices we analyse here.  In this near-field setting it is always possible to identify four nodes (a pair for each port of the device) at which to define the port currents and voltages that are related by the impedance or admittance matrices.

This lumped-device point of view is embodied in the other standard symbol for the gyrator shown in Fig.~\ref{fig:f2}~(b).  This symbol, and the current-voltage relations shown, were introduced in the seminal paper of Tellegen~\cite{TellegenPRR} reporting his invention of the gyrator concept.  Tellegen had realised~\cite{tellegenpatent1} that the lumped-element model of electric circuits was incapable of describing non-reciprocal behavior, which could readily arise in general electromagnetic theory.  He surmised that the gyrator is a minimal addition to linear network theory to make it complete, i.e., to describe any arbitrary electromagnetic linear response.  Subsequent work proved this surmise to be correct~\cite{Newcomb}.  Tellegen noted that the gyrator is a legitimate passive circuit element, neither storing nor dissipating energy.  He was hopeful~\cite{TellegenPRR, tellegenpatent1,tellegenpatent2} that a physical implementation of this device would be possible.  A partial realisation of his gyrator was achieved in subsequent investigations of the magneto electric effect~\cite{O'Dell}.  However, considering that the ideal gyrator, as he defined it, has the response Eq. (\ref{gyZ}) for {\em all frequencies}, this realisation must necessarily involve some approximation.

\begin{figure}[htb]
\centering
\includegraphics[scale=0.3]{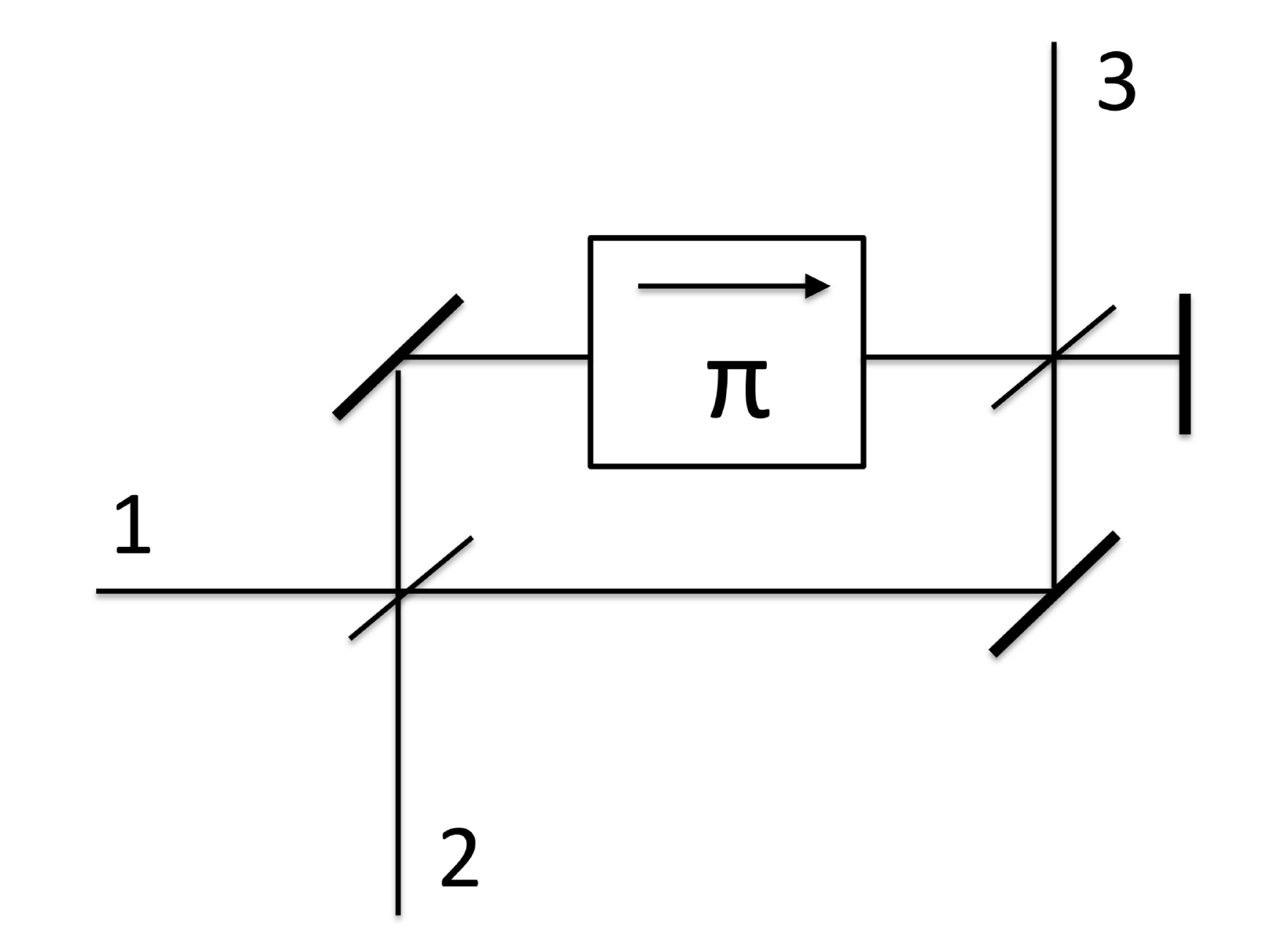}
\caption{Hogan's interferometric implementation~\cite{RevModPhys.25.253,Hoganbell} of the three-port circulator using one gyrator, using quasi-optical nomenclature.  One of the beams emerging from the Mach-Zehnder interferometer is reflected back into the structure with a mirror.  In Hogan's original discovery, magic-tees are employed rather than half-silvered mirrors (in microwave parlance, these are directional couplers).}
\label{fig:h}
\end{figure}

One clear strategy for approximate realisation is to obtain the gyrator characteristic over a limited band of frequencies.  This is what was accomplished, a few years after Tellegen's initial theoretical proposal, by Hogan~\cite{Hoganbell}.  His invention uses the non-reciprocal rotation of the polarisation of quasi-free space propagating microwaves, arising from the Faraday effect occurring in magnetised ferrite materials (see also~\cite{RevModPhys.25.253}).  A functioning gyrator was constructed that accurately approximated the ideal gyrator response in a band around a few GHz.  Hogan noted that by placing the gyrator in a interferometer structure  -- in optics language it is a Mach-Zehnder type interferometer -- ``circulation'' could be achieved (Fig.~\ref{fig:h}). 
It was immediately recognised that this circulator would have many direct applications.  Subsequent refinement of the structure, with a simplification and symmetrization of the interferometer structure, was rapidly achieved, and the Faraday circulator had achieved essentially its modern form by around 1960~\cite{LaxButton}.

In his original work Tellegen~\cite{TellegenPRR,tellegenpatent1} envisioned realisations involving non-reciprocal
~\cite{landau_electrodynamics_2009} electric or magnetic polarisations (see \cite{O'Dell}).  He did not envision a realisation based on non-reciprocal electrical {\em conduction}, but an effort was quickly made by other workers to employ this non-reciprocity -- the Hall effect -- to realise a gyrator.  We will describe this effort in Sec.~\ref{sec2} which, unlike the realisation based on the Faraday effect, ended with an apparently definitive failure.  In the present work we reexamine this failure, showing that there is an alternative approach to Hall effect gyration that is in fact successful.  It should actually be superior to the Faraday gyrator in several respects, most notably that it should permit much, much greater miniaturisation of the gyrator, and therefore of the corresponding circulator.  

The remainder of this paper will proceed as follows: Sec.~\ref{sec2} reviews the previous construction and analysis of the resistive gyrator.  We make the case in Sec. \ref{sec3} for why a reactive-coupling approach has the prospect for making a fundamentally better Hall gyrator.  The specific case of the capacitively coupled gyrator is taken up in Sec. \ref{sec:capcop}.  The extremal case of 90-degree Hall angle leads to tremendous simplifications as discussed in Sec.~\ref{goodpi}.  Our analysis is applied to a two-terminal device in Sec. \ref{twotee}, and to the four-terminal device in \ref{fourtee}, with the latter giving good gyrator characteristics.  The response of this device is significantly different if the capacitive contacts are smoothed, as analyzed in Sec. \ref{tape2}.  Three terminal devices can also lead directly to a circulator, as analysed in Secs.~\ref{sec:3T} and~\ref{tape3}.  The dual approach of inductive coupling is analysed in Sec. \ref{indgy}, where scaling arguments are given indicating why the capacitive approach is to be favoured.  A discussion of how the current ideas might be put into practice in current experimental graphene-sandwich structures is given in Sec. \ref{exgy}.  Sec. \ref{conc} gives conclusions, with some observations on the new problems posed for the quantum theory by the present device concepts.  

\section{The ``Germanium Gyrator''}\label{sec2}

At the same time as Hogan's work on the Faraday gyrator, another group of researchers (also at the Bell Telephone research laboratories) took up experiments to realise gyration using the Hall effect, and a set of results were reported employing doped germanium, motivated by the basic strategy that a low carrier density metal will exhibit a large Hall effect~\cite{Ihn}.

\begin{figure}[htp] 
\centering
\hspace{-0.1cm}
\includegraphics[scale=0.38]{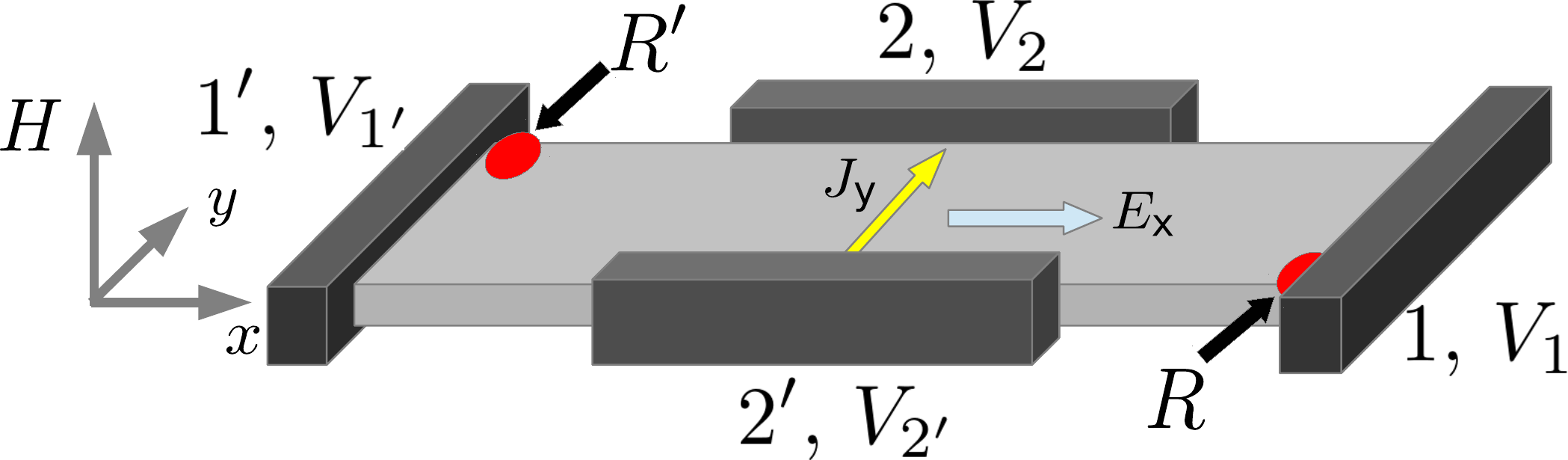}
\caption{A Hall bar geometry with four ohmic contacts, with a uniform magnetic field $H$ pointing in the $z$ direction producing a Hall effect in the electric conduction equation (\ref{HallC}) for the material.  Experiments on a thin, three-dimensional doped germanium crystal were reported in 1953~\cite{Mason1953} which attempted to realise the gyrator (four-terminal labelling corresponding to Fig. \ref{fig:f2}), with a signal field $E_x$ exciting a Hall current $J_y$.  Intrinsic inhomogeneities in the field distribution cause this device to have high losses, preventing it from successfully approximating the ideal gyrator.  For large Hall angle, the losses become concentrated at ``hot spots'' at the points $R$ and $R'$ indicated.}
\label{fig:hallbar1}
\end{figure}

Let us summarise the basic approach, following~\cite{Mason1953}.  A crystal is connected ohmically to four contacts, see Fig.~\ref{fig:hallbar1}.  The material is three dimensional but thin (the two-dimensional electron gas had not been discovered in 1953), thin enough that the conduction can be described two-dimensionally; it is assumed that there is a uniform magnetic field $H$ perpendicular to the thin conductor.  Contacts 1 and 1' (see Fig.~\ref{fig:hallbar1}), the ``current leads'' in modern parlance, will define one port of the gyrator, and contacts 2 and 2' -- the ``voltage leads'' -- will define the second port.

We consider the action of this device within the classical, Ohm-Hall framework. Here we follow the contemporary theoretical analysis of Wick~\cite{WickG} which gave an excellent accounting of the experiments performed at that time on the germanium gyrator~\cite{Mason1953}. The four contacts are equipotentials with (possibly time-dependent) potentials $V_1$, $V_{1'}$,  $V_2$, and $V_{2'}$.  It is assumed that there are no accumulations of charge inside the conductor, and that the time dynamics is quasi-static, so that the potential satisfies the two-dimensional Laplace equation
\begin{equation}
\nabla^2 V(x,y)=0.\label{laplace}
\end{equation}
The contacts then define ``Dirichlet'' boundary conditions.  The boundary conditions away from the contacts must be established by a consideration of the conduction process.   The Ohm-Hall formulation of linear electric conduction in a magnetic field is the spatially local law
\begin{equation}
-\vec\nabla V=\vec E=\rho\vec j-R_H \vec j\times\vec H.\label{HallC}
\end{equation}
The standard approximate formula for the Hall coefficient $R_H=1/(e n)$ shows why a large Hall effect is expected in a material, such as doped germanium, with a small value of the carrier density $n$.


According to Eq. (\ref{HallC}) the electric field $\vec E$ and the current density $\vec j$ are not collinear, but have a fixed angle between them, the Hall angle $\theta_H$:
\begin{equation}
\theta_H=\tan^{-1}{HR_H\over\rho}.
\end{equation}
Inverting Eq. (\ref{HallC}) and writing in componentwise form gives the matrix equation
\begin{eqnarray}
\!\!\!\!\!\!
\left(\begin{array}{l}j_x\\j_y\end{array}\right)&=&\sigma \left(\begin{array}{rr}\cos{\theta_H}&\sin{\theta_H}\\ -\sin{\theta_H}&\cos{\theta_H}\end{array}\right)\left(\begin{array}{l}E_x\\E_y\end{array}\right)\label{localC}\\
&=&\left(\begin{array}{rr}\sigma_{xx}&\sigma_{xy}\\ \sigma_{yx}&\sigma_{yy}\end{array}\right)    \left(\begin{array}{l}E_x\\E_y\end{array}\right),\,\,
\nonumber
\sigma\equiv\frac{1}{\sqrt{\rho^2+(HR_H)^2}}.
\end{eqnarray}
This equation defines the components of the conductivity tensor $\sigma_{ij}$.

With this in hand we can state the remaining boundary condition.  Away from the contacts on the boundary no currents should flow in and out of the conductor, that is,
\begin{equation}
\hat n\cdot\vec j(\vec r)=0,\,\,\,\,\vec r\in S.\label{bcj}
\end{equation} 
Here $\hat n$ is the normal vector to the boundary surface $S$.  For an ohmic conductor without a Hall effect, Eq. (\ref{bcj}) would lead to ``Neumann'' (normal derivative) boundary conditions.  However, due to the non-collinear relationship between $\nabla V$ and $j$, Eq. (\ref{localC},\ref{bcj}) imply the rotated derivative boundary condition
\begin{eqnarray}
\hat n_H\cdot\nabla V=0,\label{nH}
\end{eqnarray}
or more explicitly
\begin{eqnarray}
\cos{\theta_H}{\partial V\over\partial n}+\sin{\theta_H}{\partial V\over\partial s}=0.\label{nH2}
\end{eqnarray}
In words, this condition relates the normal derivative of $V$ to its tangential derivative (i.e., along the boundary coordinate $s$, see also Fig~\ref{fig:gy2}) in proportions determined by the Hall angle.  $\hat n_H$ is the normal unit vector rotated by (minus) the Hall angle.

\begin{figure}[htp] 
\centering
\includegraphics[scale=0.25]{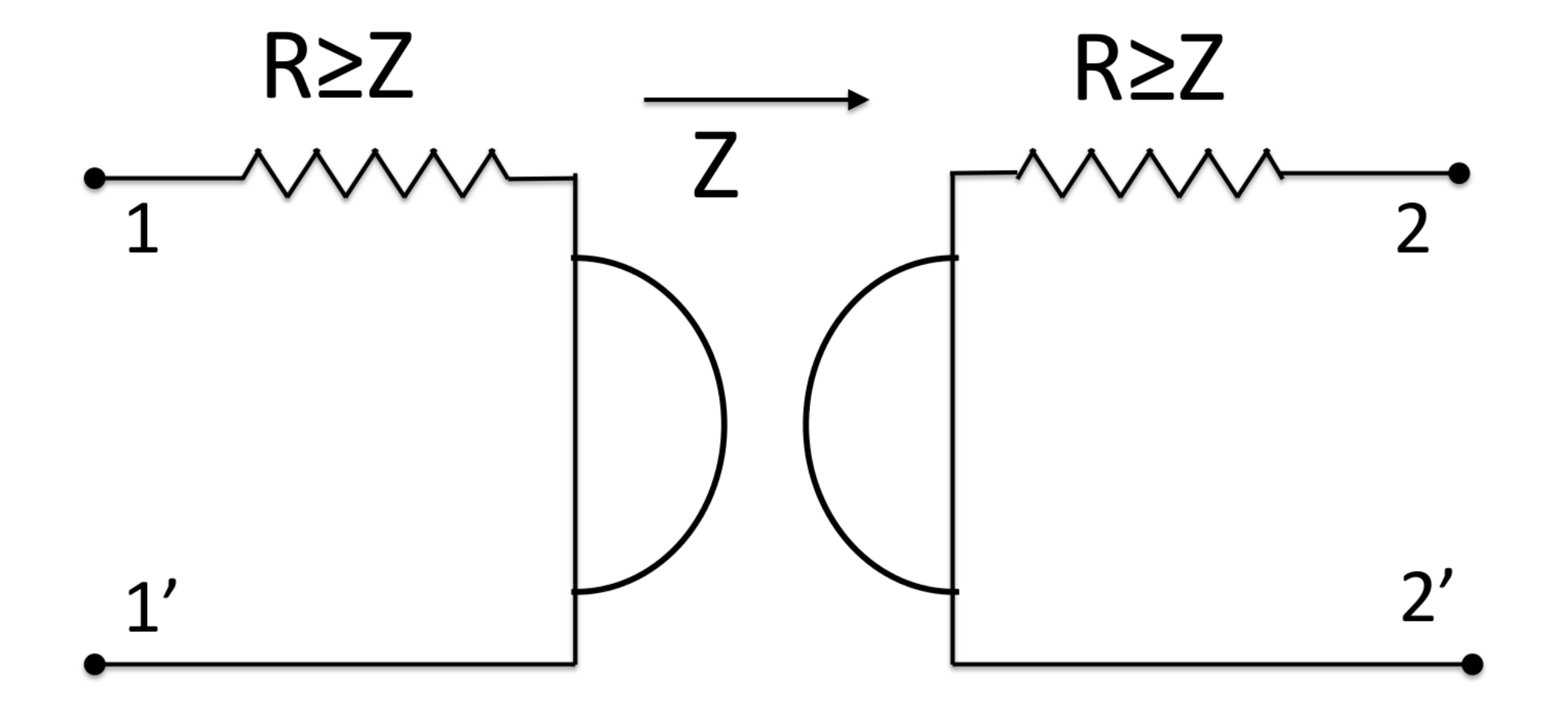}
\caption{A circuit representation of the response that is achievable using the ohmically-contacted Hall bar of Fig. \ref{fig:hallbar1}.  The lossy (diagonal) part of the impedance matrix $R$ fundamentally cannot be smaller than the antisymmetric lossless gyration resistance $Z$.}
\label{fig:lossygy}
\end{figure}

Wick~\cite{WickG} gave a very general solution to this problem, for a 2D conductor of arbitrary polygonal shape, using conformal mapping techniques; his analysis was used repeatedly in subsequent studies of such problems~\cite{RendellGirvin}, up to the present~\cite{PhysRevB.78.035416,Skachko13122010}.  His key observation, for the present purposes, is that gyration can only be poorly approximated by this device. 
He establishes his ``no go'' result with a simple argument which shows that if $I_2=I_{2'}=0$ (i.e., no total current flowing through leads 2 and $2'$), the potentials $V_2$ and $V_{2'}$ cannot lie outside the range between $V_1$ and $V_{1'}$.  This permits him to prove that there must be an input resistance $R$ at both input ports, at least as large as the gyration resistance.  Thus, the best approximation to a gyrator that can be achieved with the ``resistive gyrator'' is as shown in Fig.~\ref{fig:lossygy}. 
Wick takes pains to point out that this result is independent of the sample shape, and is also true ``in the limit of infinite magnetic field.'' By this he means the limit $\theta_H\rightarrow\pi/2$, with $\sigma$ being finite, or, in other words, 
\begin{equation}
\sigma_{xy}=-\sigma_{yx},\,\,\,\sigma_{xx}=\sigma_{yy}=0.\label{QHall}
\end{equation}
$\theta_H=\pi/2$ is the extremal case, since $\theta_H>\pi/2$ would violate the second law of thermodynamics.

While further research went on to attempt to establish useful applications for Hall gyrator~\cite{Mason,ShockleyMason,Hilbinger,Grubbs1,Grubbs2}, Wick's result indicated a fatal flaw in this approach for most purposes.  A few investigators considered the replacement of the ordinary ohmic contact with, for example, muti-contact terminals with the insertion of reactive elements\cite{Grutzmann,Arlt196075,Kroemer1,Kromer2}, to reduce resistive loss (somewhat presaging the message of the present paper).  But overall, it seems that the message of Wick's work was so well understood in the community that, when the quantum Hall regime (in which the conditions Eq. (\ref{QHall}) are satisfied) was actually observed 25 years later, it was understood without question that a two-terminal contact resistance would be present~\cite{Kawaji1978,PhysRevB.29.3749,Delahaye1993,RendellGirvin}, despite the ``dissipationless'' nature of the quantum Hall state.  B{\"u}ttiker subsequently gave a satisfactory treatment of this input resistance in the quantum theory~\cite{PhysRevB.38.9375}. The only further contemporary effort along this direction of which we are aware is the intriguing  proposal of Chklovskii and Halperin \cite{steptrans} for a step-up transformer implemented in a multiply-connected Hall-bar geometry; this proposal in fact (unwittingly) involves the cascade connection of two of the old resistive gyrators, using the transformer equivalence first noted in Tellegen's original work \cite{TellegenPRR}.  In such a transformer application, the losses identified by Wick need not be a serious limitation. 

While we tend today to view the $\theta_H\rightarrow\pi/2$ limit as profoundly quantum mechanical, in fact many details of the transport phenomenology are well captured by the Ohm-Hall classical theory in this limit.  Its treatment of boundaries can indeed be questioned; with the Hall boundary conditions a normal electric field is generally present on the insulating walls, implying a boundary layer of electric charge.  But two-dimensional electrostatics does not permit a localised line of charge in such a geometry, there is always a long-range tail of charge density into the bulk.  But this difficulty was examined in detail within the quantum treatment; MacDonald {\em et al.}~\cite{PhysRevB.28.3648} found that this boundary charge smearing in fact leads only to small quantitative errors compared with the line-charge model.  Despite all potential difficulties, the classical theory has indeed been extremely successful in giving detailed, quantitative predictions of transport behavior in suspended graphene Hall bars~\cite{PhysRevB.80.045408}.

The quantum and classical models are even in agreement on the question of {\em where} the two-terminal dissipation occurs in the quantum Hall state.  Although Wick's argument is clearly correct, it nevertheless may be viewed as paradoxical that the classical model is capable of describing any dissipation in the $\theta_H\rightarrow\pi/2$ limit.  After all, dissipation classically can be accounted for by integrating the Joule power density; using Eqs. (\ref{HallC},\ref{localC}):
\begin{equation}
P_{diss.}=\int_A\vec E\cdot\vec j\,dxdy=\sigma\cos{\theta_H}\int_A|\vec\nabla V|^2\,dxdy.\label{nodiss}
\end{equation}
Since $\cos{\theta_H}=0$, there is ``obviously'' no dissipation possible.  This argument is wrong because the fields do not have finite limiting behaviour as $\theta_H\rightarrow\pi/2$.  As reviewed clearly by Rendell and Girvin~\cite{RendellGirvin}, the fields become divergent at the ends of the ohmic contacts, on either the left or right sides according to the direction of the magnetic field (R and R' points in Fig.~\ref{fig:hallbar1} for the orientation of $H$ in the figure) depending on the sign of $\theta_H$, as  $|\theta_H|\rightarrow\pi/2$.  The fields are well behaved elsewhere; thus the argument of Eq. (\ref{nodiss}) is {\em almost} right: Joule heating goes to zero everywhere, except for ``hot spots'' (becoming Dirac delta functions, in fact) at the  $R$ and  $R'$ points.  This hot spot behaviour is observed experimentally~\cite{hotspot,Kawaji1978}, and also has a simple interpretation in a quantum treatment\cite{PhysRevB.38.9375}, where the dissipation is ascribed to a sudden change of the local chemical potential as the quantum edge states enter the lead reservoirs.  

\section{Reactive Coupling Approach}\label{sec3}

This last observation has directed the approach that we report in this paper, which analyses alternative device schemes that will achieve gyration in the ``quantum'' Hall limit $\theta_H\rightarrow\pi/2$ without accompanying two-terminal resistance.  We will confine ourselves to classical reasoning: we have argued above that the classical Ohm-Hall picture is remarkably successful in explaining the phenomenology of Hall-device conduction, and we find it an economical and insightful tool for searching for new schemes.  Quantum analyses of these schemes will certainly lead to further insights, but we will not undertake them here.

Since we see that the culprit preventing $P_{diss.}$ in Eq. (\ref{nodiss}) from being zero is a singularity arising from the incompatibility of the ohmic and insulating boundary conditions, we can investigate contactless, or reactive, means of contacting the Hall conductor.  We find both an inductive and a capacitive scheme in which the new boundary conditions avoid dangerous boundary singularities as $\theta_H\rightarrow\pi/2$.  The fields have finite limits everywhere, and the argument given above applies: as $\cos{\theta_H}\rightarrow 0$, $P_{diss.}$ goes to zero -- the ``quantum'' Hall state indeed gives a dissipationless device.  A pure gyrator is not directly achieved, but with proper choice of design excellent approximations to gyration should be achievable in convenient frequency regions, and with physical device dimensions far smaller than for the corresponding Faraday gyrator.  

While both the inductive and capacitive schemes have appealing features, we believe that the capacitive coupling scheme has the greatest potential for being realised experimentally, and has the greatest potential miniaturisability; thus we will explore this scheme in the greatest detail in the following.

\section{Capactively coupled Hall effect gyrator}\label{sec:capcop}

\begin{figure}[htp] 
\centering 
\includegraphics[scale=0.3]{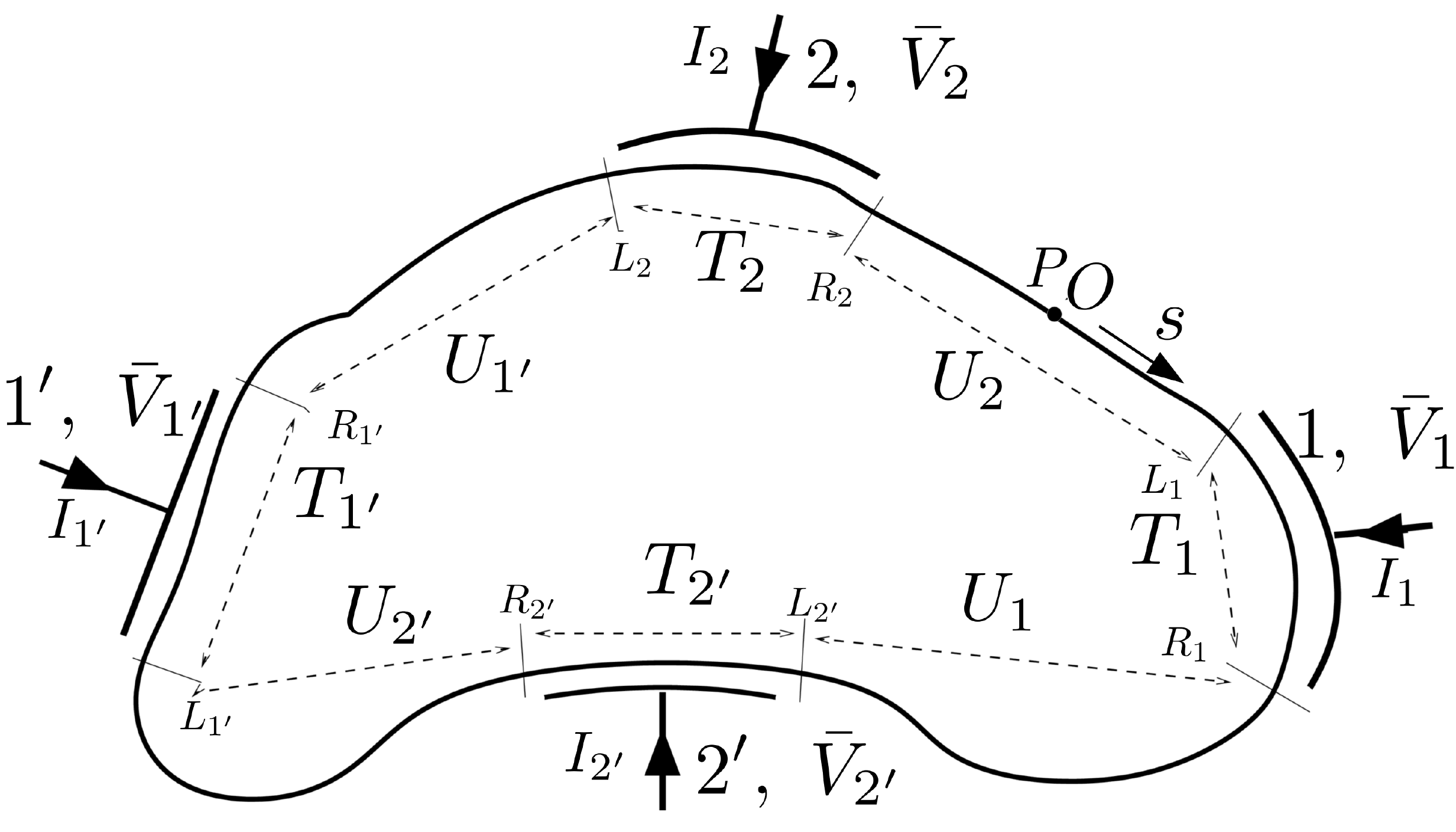}
\caption{Arrangement for four-terminal capacitive coupling to a 2D Hall conductor.  The coordinate measured along the perimeter is labeled $s$, the origin of this coordinate is labelled $O$, ending at the same point at perimeter length $P$.  Terminal segments $T$ are in capacitive contact with external electrodes at a.c. potentials $\bar V$.  The left and right $L/R$ endpoints of the $T$ segments are labeled.  $T$ segments are separated by uncontacted insulating segments $U$.  For Hall angle $\theta_H=\pi/2$ the response of this device is independent of the shape of the perimeter of the structure.}
\label{fig:gy2}
\end{figure}

We will now state a new boundary condition that is appropriate for the case of a segment of boundary of a Hall conductor forming one side of a capacitive coupling as shown in Fig.~\ref{fig:gy2}.  Such a capacitor will be characterised by having a capacitance per unit perimeter length $c(s)$. While at this point in our discussion $c(s)$ should be viewed purely as a phenomenological capacitance function, it will be important for the physical discussion given in Sec. \ref{exgy} that this function incorporate the full electrochemical capacitance to the Hall material, including the quantum capacitance~\cite{LuryiQcap}. Writing $c(s)$ this as a function of the perimeter coordinate $s$ allows the possibility that the capacitor has smoothly variable strength around the perimeter.  We will see that piecewise constant capacitances are completely reasonable, in the sense that step changes in capacitance do {\em not} lead to any singular behaviour of the fields, unlike for the case of abrupt ending of ohmic contacts.

We consider the external capacitor electrode to be a good conductor, and thus all at a single potential $\bar V$.  If at point $s$ on the perimeter the potential at the edge of the Hall conductor is $V(s)$, then the displacement current density $j_D(s)$ at that point of the capacitor, equal to the current density inside the Hall material directed normal to the edge $\hat n\cdot\vec j(s)$, is given by the ordinary capacitance equation
\begin{equation}
\hat n\cdot\vec j(s,t)=j_D(s,t)=c(s){d\over dt}(\bar V(t)-V(s,t)).
\end{equation}
The static case is uninteresting, and we have made all quantities explicit functions of time $t$.  Following Eqs. (\ref{bcj}) and (\ref{nH}), the normal current is proportional to the rotated projection of the field gradient:
\begin{equation}
\hat n\cdot\vec j(s,t)=-\sigma\,\hat n_H\cdot\nabla V(s,t)=c(s){d\over dt}(\bar V(t)-V(s,t)).
\end{equation}
We may write this equation in the frequency domain, giving our final boundary-condition equation:
\begin{equation}
-\sigma\,\hat n_H\cdot\nabla V(s,\omega)=i\omega c(s)(\bar V(\omega)-V(s,\omega)).\label{ourB} 
\end{equation}
While this is a perfectly well-posed mixed, inhomogeneous boundary condition for the Lapace equation, we are not aware that it has been previously examined.  It is applicable around the entire boundary, as the regular insulating boundary condition for the Hall conductor (Eqs. (\ref{nH},\ref{nH2}) above) corresponds to a region of boundary with $c(s)=0$.  Ohmic boundary conditions are in some sense treated by taking $c(s)\rightarrow\infty$, but this case will not come up in the following, and by keeping $c(s)$ finite we avoid the singular behaviour of the fields discussed above.  Note that our boundary condition (\ref{ourB}) is complex-valued; this has the normal interpretation for a.c. electrical problems, that the real part of the field is the in-phase response and the imaginary part of the field is the out-of-phase or quadrature response; that is, the when driven with a field at frequency $\omega$, the temporal response is ${\rm Re}(V(r,\omega))\cos{\omega t}+{\rm Im}(V(r,\omega))\sin{\omega t}$.  While we will apply these boundary conditions for $\omega$ up to microwave frequencies, we will consider only cases where the device dimensions are much smaller than the wavelength of radiation at these frequencies; in this near field limit the quasi-static analysis of the bulk conduction as determined by the Laplace equation Eq. (\ref{laplace}) still applies.  

The boundary conditions for the ohmic and capacitive contacts (Eqs.~(\ref{nH},\ref{nH2}) and~\eqref{ourB}) have different behavior under conformal transformation. The ohmic case is conformally invariant since it only fixes the direction of $\vec E$ with respect to the boundaries. Instead, Eq.~\eqref{ourB} is a condition on the values of $\vec E$ along a direction, hence it is not conformally invariant. 
Therefore the conformal mapping methods~\cite{MorseF,WickG,PhysRevB.78.035416} cannot easily be applied to the electrostatic problem in our case; however, as we will show shortly, we can calculate all device quantities of interest (analytically for $\theta_H=90^\circ$) without resort to conformal mapping techniques.
   
The problem of finding the two-port response, e.g., the admittance matrix $Y$ (or $Z$ or $S$), is now straightforwardly posed: given a geometry as in Fig.~\ref{fig:gy2}, we identify four terminal segments $T_{1,1'}$ and $T_{2,2'}$ and four uncontacted, insulating segments $U_{1,1'}$ and $U_{2,2'}$.  There is no capacitance in the $U$ regions.  $c(s)$ will be nonzero along the $T$ segments; we will analyse both the case of constant capacitance per unit length, and the case where the capacitance goes smoothly to zero at the ends of these segments.  Smoothing will cause significant differences in the response, but this difference does not modify the main features in the relevant frequency range. The a.c. terminal potentials $\bar V_{i=1,1',2,2'}$ will, as stated above, be taken as constants (possibly complex) in each of the terminal segments $T_i$.  Solving the field problem as a function of $\omega$ gives normal boundary currents $\hat n\cdot\vec j(s,\omega)=-\sigma\,\hat n_H\cdot\nabla V(s,\omega)$; integrating gives the terminal currents
\begin{equation}
I_i(\omega)=\int_{T_i}\hat n\cdot\vec j(s,\omega)ds=-\sigma\int_{T_i}\hat n_H\cdot\nabla V(s,t)ds.\label{curcal}
\end{equation}
These terminal currents are linear functions of the terminal potentials:
\begin{equation}
I_i(\omega)=\sum_{j=1,1',2,2'} y_{ij}\bar V_j.
\end{equation}
The coefficients in this equation are admittances, but one further calculation is needed to obtain the two-port admittance matrix $Y$ from them.  We must enforce the condition that the terminal pairs $T_{1}-T_{1'}$ and $T_{2}-T_{2'}$ act as ports.  The pair $T_i-T_j$ is a port if $I_i=-I_j$.  One must determine the relative potential between our two ports, as measured by, e.g., $\bar V_{1}-\bar V_{2}$, which will cause the port condition $I_{1}=-I_{1'}$ to be satisfied; then the other condition $I_{2}=-I_{2'}$ is automatically satisfied, since the total current entering the Hall conductor is zero.  Then the port currents are functions of the port voltages, viz.,
\begin{eqnarray}
I_1&=&Y_{11}(\bar V_1-\bar V_{1'})+Y_{12}(\bar V_2-\bar V_{2'}),\nonumber\\
I_2&=&Y_{21}(\bar V_1-\bar V_{1'})+Y_{22}(\bar V_2-\bar V_{2'}),
\end{eqnarray}
thus defining the $2\times 2$ port admittance matrix $Y$.  We will now investigate under what conditions the gyrator matrix Eq. (\ref{gyY}) is obtained.

One further comment about going from terminal to port response: in the electrical literature, it is often assumed without discussion~\cite{Newcomb} that the ports are electrically isolated, meaning that there is identically vanishing dependence on the potential difference between the two ports ($\bar V_1-\bar V_2$ in the analysis above).  Under these circumstances the port current condition is also automatically satisfied.  This isolation is not present in our device 
(e.g., current can, in principle, flow from terminal 1 to 2).  It is understood that in many circumstances this input-output isolation is not necessary for proper functioning of the device; if it is needed, it can be achieved by separate isolation (e.g., transformer coupling).  This issue will arise one further time in the present paper, in the analysis of the three-terminal gyrator in Sec.~\ref{sec:3T}.

\subsection{Response requires only boundary calculation for Hall angle $\pi/2$}\label{goodpi}

\begin{figure}
        \centering
        \hspace{-0.02\textwidth}
                 \includegraphics[scale=.510]{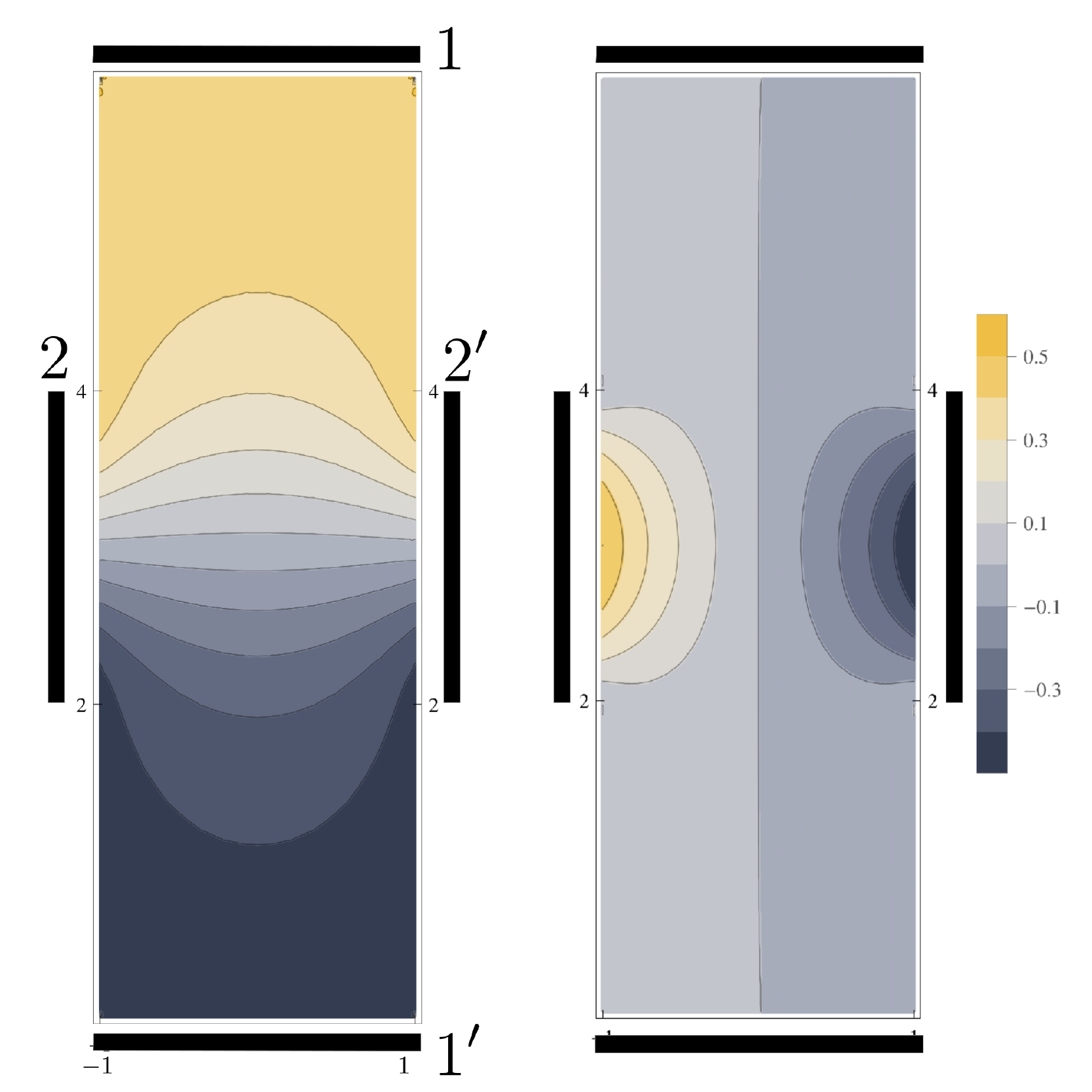}
        \caption{Two dimensional plots of the in-phase (left panel) and out-of-phase (right panel) potential fields for a capacitively coupled four-terminal Hall-bar device, for $\theta_H=\pi/2$.  The frequency of the applied field is $\nu_{gy}=\sigma/2C_L$ (Eq. (\ref{pgy2})), the first perfect gyration frequency.  Contacts span the entire (length=2) of the top and bottom edge of the bar, with $\bar V=\pm.5V$. The position of the contacts are indicated with the black bar.  Length-2 contacts are centered on the left and right of the bar, with $\bar V=0$.  The capacitance function $c(s)$ is a constant on all the terminal boundaries.  Perfect gyration requires that there be current only from the left to the right terminal, and it be in-phase.  We see that the in-phase current density (left panel), which follows the potential contours in accordance with the guiding-center principle, does indeed flow smoothly from one contact to the other.  Note that these contacts are not equipotentials, as they would be for ohmic contacts; at this perfect gyration frequency, however, the top and bottom terminals are at equipotentials.  There is a non-zero out of phase current density flow (right panel), but it is purely local to the contact, and results in no net current.}\label{fig:vgim}
\end{figure}

We now proceed to explicit calculations of several Hall gyrator structures.  Since, as we confirm shortly, lossless operation is achieved in the case of Hall angle equal to its extremal value of $\pi/2$, and so $\theta_H=\pi/2$ will be the principal focus of our study.  We will see that our calculations are well behaved at this value, so that no subtle limits need to be taken.  However, we find that the field equations simplify remarkably for $\theta_H=\pi/2$, permitting closed-form solutions for a wide class of device structures.  This simplification arises from examining our boundary condition equation (\ref{ourB}) for this case; recalling Eq. (\ref{nH2}), we obtain
\begin{equation}
-\sigma\,{\partial V(s,\omega)\over\partial s}=i\omega c(s)(\bar V(\omega)-V(s,\omega)).\label{ourqB} 
\end{equation}
This boundary condition equation now contains only the tangential derivative of the potential, which involves only potential values at the boundary.  Thus, this equation is now a closed one-dimensional condition in the boundary coordinate $s$, fully determining the field on the boundary without reference to the interior of the conductor.  The field in the interior of the conductor is still well defined, but is entirely a slave of the boundary potential; the full field can be calculated by considering the perimeter field as a Dirichlet boundary condition.  But all device-response coefficients are purely functions of the perimeter field, so the interior field need never be calculated. Two dimensional plots of the in-phase and out-phase fields $V(s,\omega)$, for the Hall bar with four capacitive contacts (the same setup as in Fig.~\ref{fig:hallbar1})  is shown in the Fig.~\ref{fig:vgim} for a frequency for which perfect gyration occurs.

Note that the calculation of the terminal currents also takes a much simpler form in this case; Eq. (\ref{curcal}) becomes
\begin{align}
I_i(\omega)&=-\sigma\int_{T_i}{\partial V(s,\omega)\over\partial s}ds
\nonumber\\
&=\sigma(V(s=L_i,\omega)-V(s=R_i,\omega)).\label{curcalH}
\end{align}
Thus, the current is simply given by the difference of the potential from the left point of the capacitor $L_i$ to the right point $R_i$  (see also Fig.~\ref{fig:gy2})~\footnote{There is a remarkable resemblance between the purely classical Eq (\ref{curcalH}) and the quantum relation between edge potential and edge state currents in the quantum picture of the Hall bar \cite{PhysRevB.38.9375}, this is clearly visible in the two terminal case.}. Furthermore, the field solution is completely independent of the shape of the boundary; it can be deformed at will (as suggested by Fig. \ref{fig:gy2}), and the boundary potential and all device response coefficients will be unchanged as long as the perimeter length $P$ and the capacitance function $c(s)$ are unchanged.  Note that for general boundary conditions, the solution on the perimeter can be written as an integral over the perimeter (cf. Eq. (7.2.12) of Ref. \cite{MorseF}); but the kernel of this integral is a Green function which, in the general case, is globally sensitive to the detailed structure of the entire conductor.  Thus, our situation is quite special.

It is valuable to note that the homogeneous part of Eq. (\ref{ourqB}) is a one-dimensional Dirac eigenvalue equation, with periodic boundary conditions from 0 to $P$, with $c(s)$ playing the role of the position-dependent mass of that Dirac equation.  The two-component Dirac spinor consists of the real and imaginary part of $V$.  The eigenfrequencies $\omega_n$ of this equation are equally spaced:
\begin{equation}\label{eq:plasmonsw}
\omega_n={2n\pi\sigma\over\int_0^Pc(s)ds}.
\end{equation}
We will see that these eigenmodes have the physical meaning of chiral edge magnetoplasmons of the Hall conductor; they will set the scale of frequency at which interesting gyrator behavior occurs.

Magnetoplasmons have been investigated thoroughly in 2D Hall conductors~\cite{PhysRevLett.54.1710,PhysRevLett.54.1706}, including in the quantum Hall regime~\cite{Gorkov,Andrei}. In Sec.~\ref{exgy} we will discuss details of how this work has developed up to the present, and what suggestions it makes for the physical implementation of the devices analysed here.

\subsection{Two-terminal device}\label{twotee}

While it has no application for gyration (two terminal devices must be reciprocal), the solution to the simple two-terminal problem is instructive, especially for the insight that it gives into the edge magnetoplasmon dynamics in this system.  We consider the special case of the two capacitors with constant capacitance per unit length attached to the Hall conductor.  Suppose the length of the capacitor is $L$ and the capacitance per unit length is $c$, so that the total lead capacitance is $C_L=c L$.  Then the (scalar) admittance of the device is calculated to be
\begin{equation}
Y(\omega)=i\sigma\tan{{\omega C_L\over 2\sigma}}.\label{2ad}
\end{equation}
This solution is still correct for the case when the two leads have different widths, but the $c$-$L$ products should be the same.  The placement of the leads around the perimeter is arbitrary, the lengths of the insulating regions between the leads are irrelevant, and the conductor can be of arbitrary shape (including sharp turns).  Interior holes in the conductor also play no role.  Note that the poles of this admittance coincide with magnetoplasmon eigenfrequencies (as defined in Eq.~\eqref{eq:plasmonsw}), and that the low-frequency limit, $i\omega C_{L}/2$, is that of two capacitors $C_L$ in series. 

In fact, the admittance function (\ref{2ad}) is a familiar one.  It is identical to that of a segment of transmission line with characteristic impedance $1/\sigma$ and transit time (wave velocity times length) of $\tau=C_L/2\sigma$, open-circuited at the end.  An important feature of this transmission-line response is that a short voltage pulse applied to it is perfectly reflecting, but with a transit-time delay of $2\tau$.  When this pulse is applied to the two-terminal Hall device, where does the pulse live during this $2\tau$ transit time?  The answer is that when the pulse arrives at the capacitor electrodes, it produces a non-zero field in the Hall conductor only in the immediate vicinity of the right edge (points $R_i$) of the two capacitors.  This localised edge field propagates, in a dispersionless way, counter-clockwise around the edge of the conductor, with velocity
\begin{equation}
v_{pl}=L/2\tau=\sigma/ c.\label{vel}
\end{equation}
After time $2\tau$ these two edge excitations reach the left end (points $L_i$) of the capacitors, causing a re-emission of the radiation pulse back into the leads.  For normal device parameters, this plasmon propagation velocity is far smaller than the speed of light; thus, this ``simulated transmission line'' is very compact compared with the corresponding real transmission line.   

\subsection{Four terminal device: the gyrator}\label{fourtee}

We now return to the four terminal device, revisiting the approach of Mason {\em et al.} \cite{Mason1953} (Fig. \ref{fig:hallbar1}), with ohmic contacts replaced by capacitive contacts.  For the case of uniform capacitance, $c(s)=$const., and four equal contact capacitors with capacitance $C_L$, the exact solution for the two-port response matrices is elementary.  The admittance is
\begin{equation}
Y_{2P}(\omega)={\sigma\over 2}\left(\begin{array}{cc}i\tan{\omega C_L\over\sigma}&-1+\sec{\omega C_L\over\sigma}\\ 1-\sec{\omega C_L\over\sigma}&i\tan{\omega C_L\over\sigma}\end{array}\right),\label{Y2P}
\end{equation}
which when inverted gives the two-port impedance
\begin{equation}
Z_{2P}(\omega)={1\over\sigma}\left(\begin{array}{cc}-i\cot{\omega C_L\over 2\sigma}&-1\\ 1&-i\cot{\omega C_L\over 2\sigma}\end{array}\right).\label{zeel}
\end{equation}
Note that $Y$ and $Z$ satisfy the conditions for multiport lossless response, which are~\cite{Newcomb} that the imaginary part of the matrix be symmetric and an odd function of frequency, while the real part is antisymmetric and an even function of frequency.  (These conditions are also trivially satisfied for the one-port device in Eq. (\ref{2ad})). This condition is equivalent to $P_{diss.}=0$ (see Eq.~\eqref{nodiss}).  This confirms that the argument given using Eq. (\ref{nodiss}) applies to our calculation, given that there is no singularity involved in going to the $\theta_H\rightarrow\pi/2$ limit.  The presence of a nonzero antisymmetric part indicates the non-reciprocal response of the device.   The periodicity of the response coefficients is again indicative of ``delay-line'' behavior; an equivalent circuit for this response is that of a gyrator with series lossless transmission-line stubs at the inputs, replacing the resistors of Fig. \ref{fig:lossygy}.  We will not explore here the details of the edge-magnetoplasmon propagation that causes this multi-port response.  

\begin{figure*}[htp]
\centering
\includegraphics[scale=1.3]{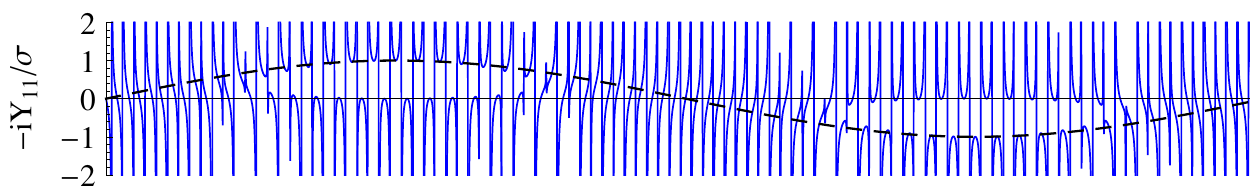}\\
\includegraphics[scale=1.3]{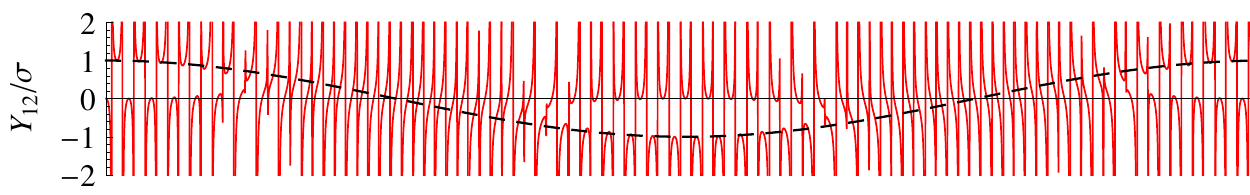}\\
\includegraphics[scale=1.3]{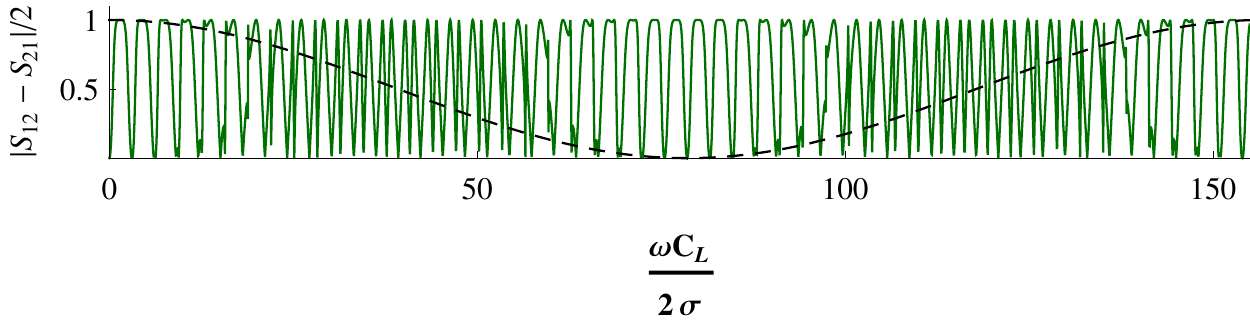}
\caption{The behavior of the admittance (Eq. (\ref{Y2P})) and of the scattering parameters (cf. Eqs. (\ref{gyZ},\ref{gyY}) with $Z_0=1/\sigma$) for the two port device with smoothed capacitive contacts.  The capacitance function is as given in Eq. (\ref{taper}) with  $\lambda=L/12$. We show a large range of frequency, including about 100 poles of the admittance, and about 50 good gyration points as given by Eq. (\ref{pgy}), in order that the slow modulation of $\omega$ on the scale $\sigma/c\lambda$ can be seen.  The dashed lines showing this modulation sinusoid are guides to the eye.  Top panel: (11) component (pure imaginary) of the admittance matrix $Y_{2P}$. Middle panel: (12) component of the admittance matrix $Y_{2P}$ . Bottom plot: $\vert (S_{2P,\lambda})_{1,2}-(S_{2P,\lambda})_{2,1}\vert/2$.  Due to unitarity, this quantity can only attain the value unity if good gyration is achieved ($S$ matrix proportional to Eq. (\ref{gyS})).  We see that despite the modulation caused by smoothing, perfect gyration occurs regularly along the frequency axis, at points close to those given by the formula Eq. (\ref{pgy}) for the unrounded case.
\label{fig:testf}
}
\end{figure*}

\begin{figure}[htp]
\centering
\includegraphics[scale=0.6]{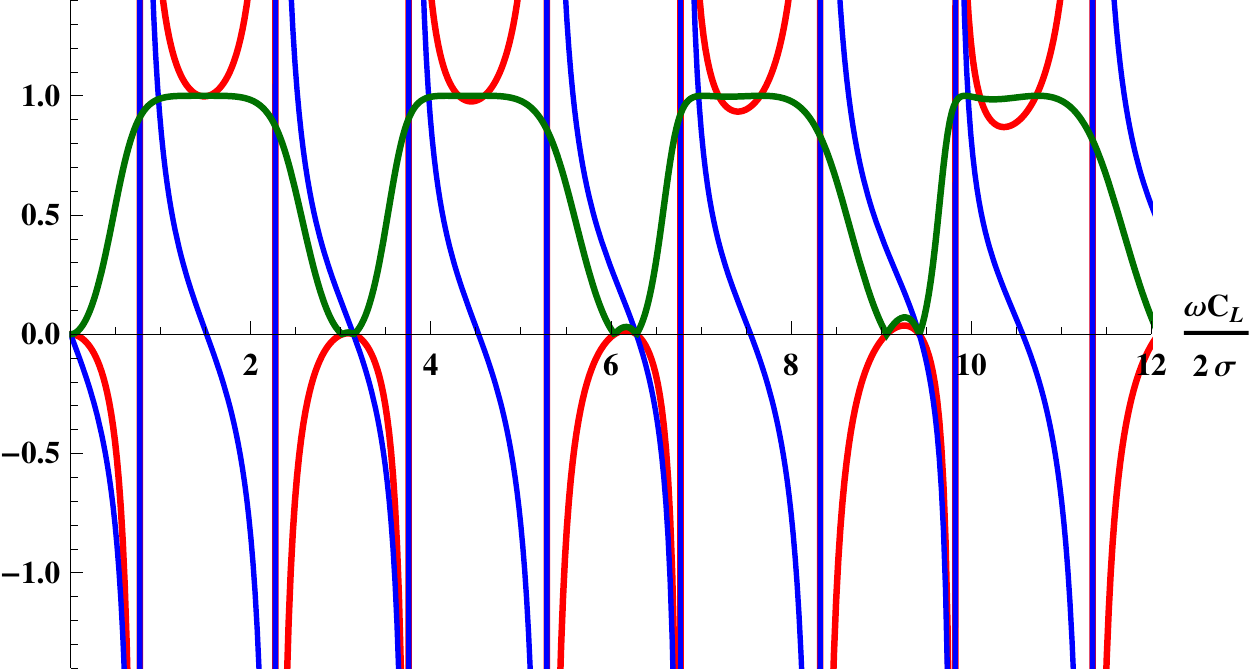}\\
\caption{Close up of Fig.~\ref{fig:testf}:  $\vert (S_{2P,\lambda})_{1,2}-(S_{2P,\lambda})_{2,1}\vert/2$, $\vert (Y_{2P,\lambda})_{11}\vert$ and $\vert (Y_{2P,\lambda})_{21}\vert$ in green, blue and red respectively, at low frequency. Over the first period of the response (i.e., for $0<\omega C_L/2\sigma<\pi$) the response shown here is almost indistinguishable from that of the constant capacitance, Eqs. (\ref{Y2P},\ref{Y2Pl}). }
\label{fig:syy2p}
\end{figure}

While the response for $\theta_H=\pi/2$ is thus manifestly lossless, we can further make a perturbation argument for the case of Hall angle slightly smaller than 90 degrees, $\theta_H=\pi/2-\epsilon$: to lowest order in $\epsilon$ the field solutions can be taken to be independent of $\epsilon$, so that the total dissipation of the device will, using Eq. (\ref{nodiss}), be proportional to $\epsilon$, but with ``nonuniversal'' coefficients (i.e., dependent on the details of the device geometry).

The scattering matrix is obtained using the formula $S=(Z+Z_0 I)^{-1}(Z-Z_0 I)$~\cite{Pozar}  with $Z_0=1/\sigma$:
\begin{subequations}
\begin{eqnarray}
S_{2P}(\omega)&&=2d^{-1}
\left(\begin{array}{cc}
\cos ^2(\frac{\omega C_L}{2\sigma} )&2\sin ^2(\frac{\omega C_L}{2\sigma} )\label{sube}\\
-2\sin ^2(\frac{\omega C_L}{2\sigma} )&\cos ^2(\frac{\omega C_L}{2\sigma} )
\end{array}\right)\\
d&&={3 \cos (\frac{\omega C_L}{\sigma})-2 i \sin (\frac{\omega C_L}{\sigma})-1}
\end{eqnarray}
\end{subequations}
Note that $S_{2P}$ is unitary (because the device is lossless) and non-symmetric (because it is non-reciprocal).

The relation to gyration is especially easy to see using the impedance matrix Eq. (\ref{zeel}).  There is a series of frequencies at which perfect gyration (Eq. (\ref{gyZ})) is achieved, given by the equation $\cot{\omega C_L\over 2\sigma}=0$; these frequencies are
\begin{eqnarray}
\omega_{gy}&=&{\pi\sigma\over C_L}(1+2n),\,\,\,n\ge 0\label{pgy}\\
\nu_{gy}&=&{\sigma\over 2C_L}(1+2n)\,\,[{\mbox{Hz}}]\,\,\,n\ge 0\label{pgy2}
\end{eqnarray}
These perfect-gyration frequencies lie halfway between the plasmonic poles present in $Y$, cf. Fig. \ref{fig:syy2p}.  Thus the gyration is fundamentally non-resonant, and is present to good approximation over relatively wide ranges of frequency.

\subsection{Smoothed capacitances: two-port case}\label{tape2}

While many aspects of the device geometry are irrelevant for the port response, the details of the edge-capacitance function $c(s)$ do matter.  We now study the device behavior if the capacitors are smoothed, so that $c(s)$ goes continuously to zero at the edges of the capacitors.  Closed-form expressions of the solution of Eq. (\ref{ourqB}) are obtainable for many different tapering functions; a convenient analytic form is

\begin{equation}
c(s)=\left \{
\begin{array}{ll}
c& \vert s \vert <\frac{L}{2}\\
c\,{\rm sech}^2\left(\frac{2\vert s \vert-L}{2\lambda}\right)& \vert s \vert >\frac{L}{2}
\end{array}
\right.\label{taper}
\end{equation}
Assuming that the insulating region between contacts is many $\lambda$s long so that these is negligible overlap between these capacitance functions, we find the two-port admittance to be
\begin{eqnarray}
Y_{2P,\lambda}&=& \\
 \frac{\sigma\sin \left(\frac{L c\omega }{2 \sigma }\right)}
{\cos \left(\frac{(2 \lambda +L)c\omega}{\sigma }\right)}
&&\!\!\!\!\!\!\!\! \left(
\begin{array}{cc}
  -i \cos (\frac{(L+2\lambda ) c\omega}{2 \sigma} ) &
\sin (\frac{(L+2\lambda ) c\omega}{2 \sigma}) \\
-\sin (\frac{(L+2\lambda )c \omega}{2 \sigma} ) & -i \cos (\frac{(L+2\lambda ) c\omega}{2 \sigma})
\end{array}
\right).\nonumber\label{Y2Pl}
\end{eqnarray}
We see that for small rounding, this response exhibits a slow modulation in frequency (on the scale of $\omega=\frac{\sigma}{c\lambda}$), with the low-frequency behavior matching the unrounded ($\lambda=0$) response calculated above.  Fig.~\ref{fig:testf} plots this response and the relevant component of the scattering matrix $S_{2P,\lambda}$ or simply $S$ for slightly rounded capacitances.  As Fig.~\ref{fig:syy2p} shows, at low frequency perfect gyration occurs at the regularly-spaced frequencies as indicated by Eqs. (\ref{pgy},\ref{pgy2}).  This is indicated by $|S_{12}-S_{21}|$ attaining the value 2; because of unitarity, this can only occur if $S_{11}=S_{22}=0$ and $S_{12}=-S_{21}$, the conditions for a perfect gyrator.  We see in fact that at low frequency this condition is satisfied over wide bands.  As the modulation due to the rounded contacts begins to have an effect, the frequency dependence of $|S_{12}-S_{21}|$ is modified, but it still returns to the ideal value of 2 frequently, albeit over narrower frequency ranges.  

We note that this modulation causes the perfect gyration points to change from having a real-valued $S$ matrix ($S_{12}$, $S_{21}=\pm 1$) to being complex valued; the $S$ matrix acquires an overall reciprocal phase factor.  Since this is what one obtains for a perfect gyrator with a change of reference plane~\cite{Pozar}, it is fair to refer to this still as perfect gyration.  It would, however, require some reconsideration of the Hogan construction Fig.~\ref{fig:h}; the reference arm of this interferometer would have to have a corresponding phase change, which may cause it to be a significant fraction of a wavelength in size.  To make this construction compact in this case, the reference arm phase delay could be simulated by two cascaded Hall effect gyrators, chosen to given the correct overall net reciprocal phase~\cite{TellegenPRR}. 

Finally, we note that we have used a source impedance $Z_0=1/\sigma$ for calculating $S$.  We find that for $Z_0>1/\sigma$ the perfect gyration condition  $|S_{12}-S_{21}|=2$ continues to be satisfied for a regularly spaced set of frequencies.  For $Z_0<1/\sigma$ (the more likely case, see the discussion in Sec.~\ref{indgy}) perfect gyration no longer occurs at low frequencies; but with finite rounding, at higher frequency perfect gyration again occurs.  However, for large impedance mismatch $Z_0<<1/\sigma$ good gyration occurs only over very narrow ranges of frequency.

\subsection{Three-terminal device and the Carlin construction}\label{sec:3T}

\begin{figure}[t] 
\centering 
\includegraphics[scale=0.35]{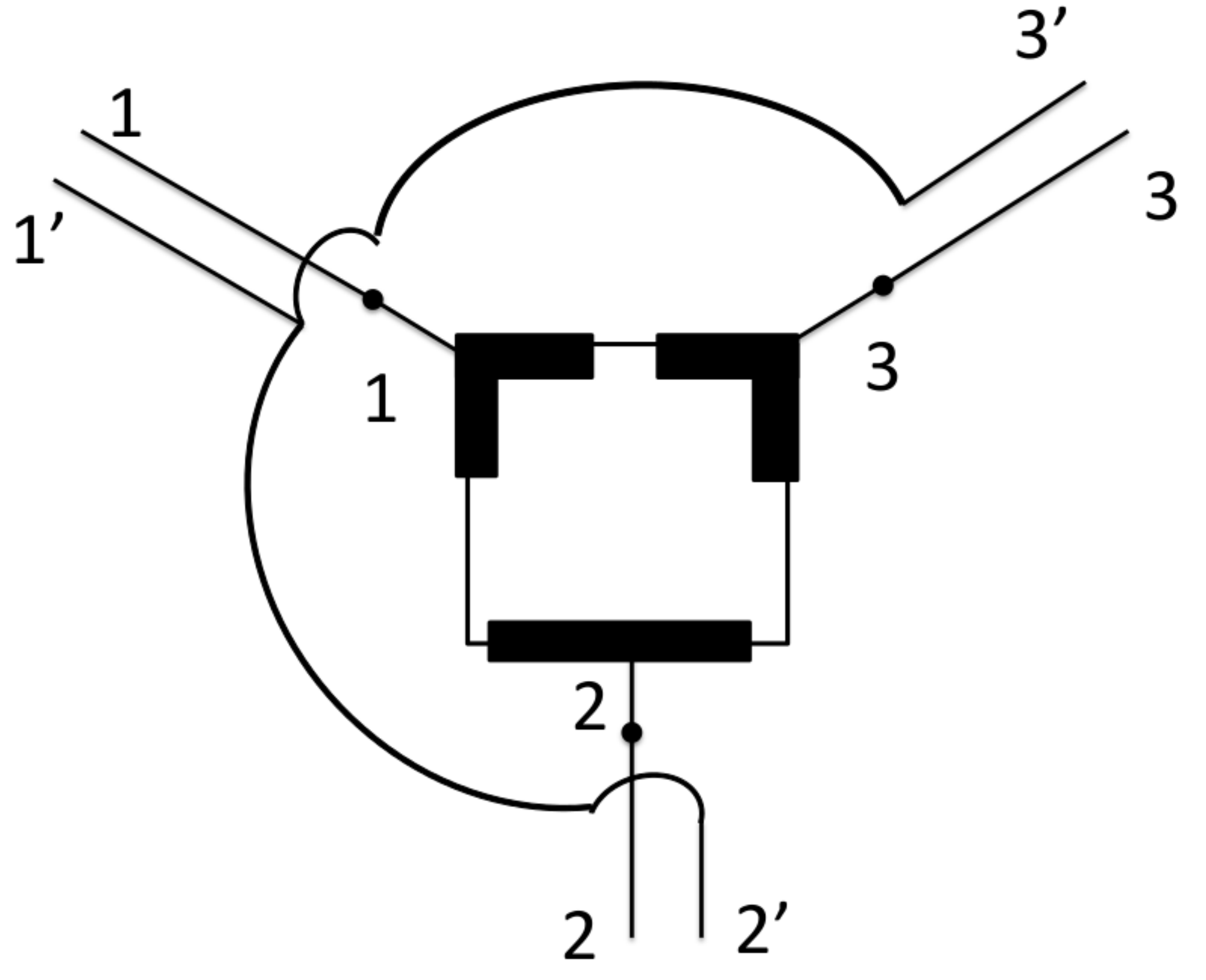}
\caption{The construction of Carlin~\cite{Carlin,Carlinbook,Newcomb} for realising a circulator of Fig.~\ref{fig:f1} and Eq. (\ref{theS}), given a nonreciprocal three-terminal device (terminals defined at solid dots) with admittance as in Eq. (\ref{3T}). Anticipating Sec. \ref{exgy} and Fig. \ref{fig:layers}, we depict the capacitive contacts as strips overlapping the edge of a rectangular piece of Hall material; the capacitances should be the same, and may or may not be rounded.  Note that the primed terminals of each of the three external ports are tied together, but should be kept away from the Hall device so that they have no ohmic or capacitive contact to it.
}
\label{fig:ccarlin1}
\end{figure}

\begin{figure}[t] 
\centering 
\includegraphics[scale=0.35]{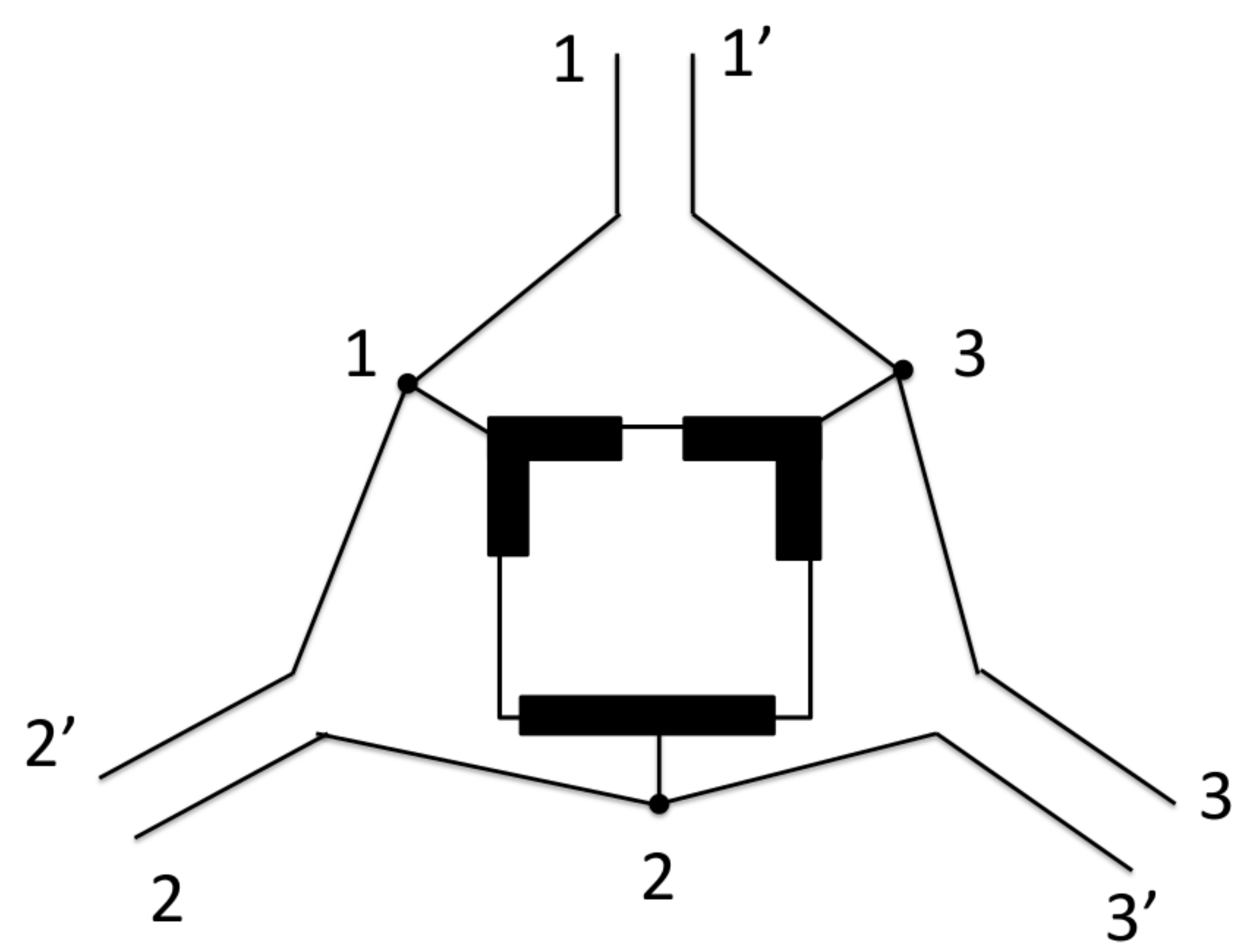}
\caption{A dual construction of Carlin~\cite{Carlin,Carlinbook} for realising a circulator with the same three-terminal device as in Fig. \ref{fig:ccarlin1}.  Phase-inverted circulation is achieved, i.e., the $S$ matrix is the negative of Eq. (\ref{theS}).}
\label{fig:ccarlin2}
\end{figure}

Carlin~\cite{Carlin,Carlinbook} (see also~\cite{Newcomb}) noted that there are several alternatives to the Hogan construction (Fig. \ref{fig:h}) for realizing a three-port circulator using a gyrator.  They are arguably more direct in that they do not require an interferometer.  In Carlin's original construction he employs the classic Tellegen gyrator (Fig. \ref{fig:f2}(b)) tied to a common ground, i.e., with terminals 2 and 2' short circuited.  This approach cannot be applied directly to our four-terminal Hall gyrator, because of the lack of input-output isolation mentioned above 
(Sec.~\ref{sec:capcop}).  However, Carlin's construction can be stated more directly: a three-terminal device with the right non-reciprocal admittance matrix (see Eq. (\ref{3T})) can be converted to a circulator with either of the two Carlin constructions Figs.~\ref{fig:ccarlin1}, \ref{fig:ccarlin2}.  The ``dual'' construction of Fig.~\ref{fig:ccarlin2} actually gives a phase-inverting circulator, with $S$ being the negative of Eq. (\ref{theS}); we are not aware of any current application of the circulator in which the phase of $S$ is relevant.


Perfect Carlin circulation is obtained with a three-terminal device with the admittance matrix
\begin{eqnarray}
Y_{3T}(\omega)&=&\left(\begin{array}{ccc}ia&b&-b^*\\-b^*&ia&b\\b&-b^*&ia\end{array}\right)\label{3T}
\end{eqnarray}
when $a=0$, $\im(b)=0$, and $\re(b)=1/Z_0$, the source impedance.  This is obtained by using a three terminal Hall device with equal contact capacitances $C_L$.  (The naive procedure of short circuiting 2 and 2' in the four-terminal device would lead to one contact effectively having capacitance $2C_L$.)  For the case of constant capacitance functions, the response is as in Eq. (\ref{3T}), with  
\begin{eqnarray}
a&=&{2\sigma\sin{\omega C_L\over\sigma}\over 1+2\cos{\omega C_L\over\sigma}}\\
b&=&\sigma{-1+\exp{-i\omega C_L\over\sigma}\over 1+2\cos{\omega C_L\over\sigma}}
\end{eqnarray}
If the device is matched ($Z_0=1/\sigma$) perfect circulation is obtained at the frequencies 
\begin{equation}
\nu_{circ}={\sigma\over 2C_L}(1+2 n)\,\,[{\mbox{Hz}}],\,\,\,\,n=1,2,...
\end{equation}

\subsection{Rounded capacitances: three-terminal case}\label{tape3}

This response matrix can also be easily calculated in the case of rounded capacitance, Eq. (\ref{taper}).  The result is
\begin{figure}[htp]
\centering 
\includegraphics[scale=0.6]{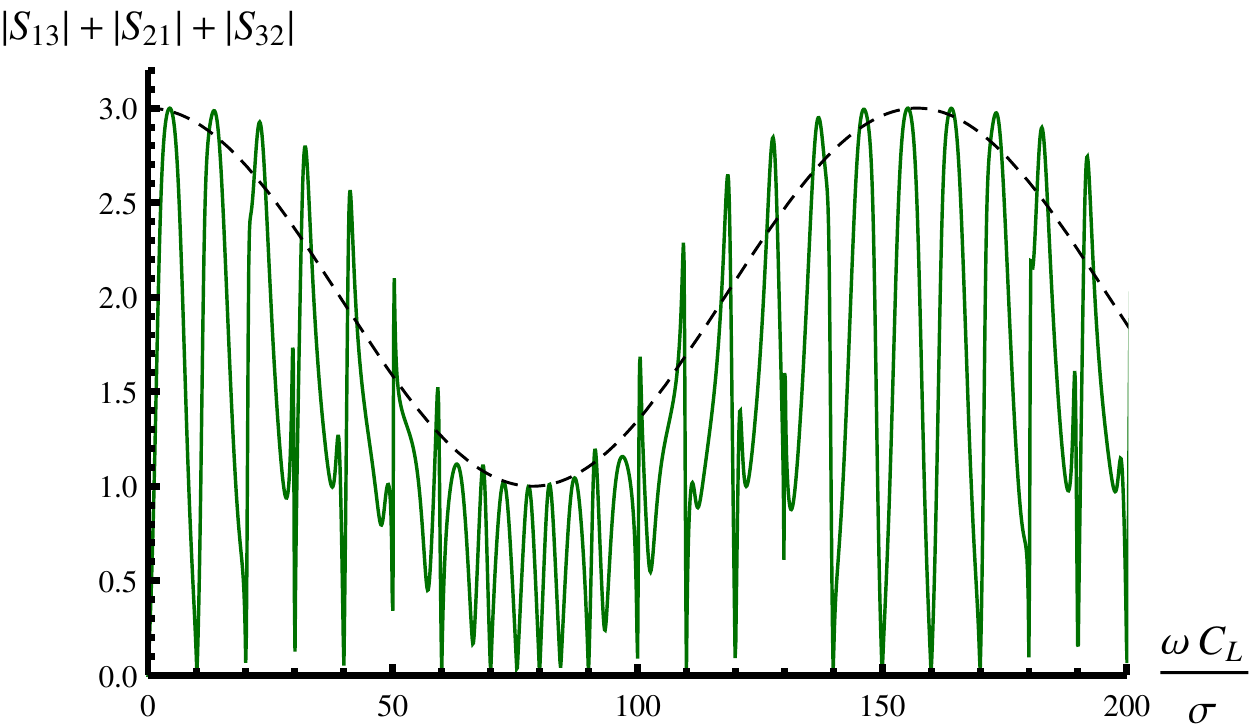}\\
\includegraphics[scale=0.6]{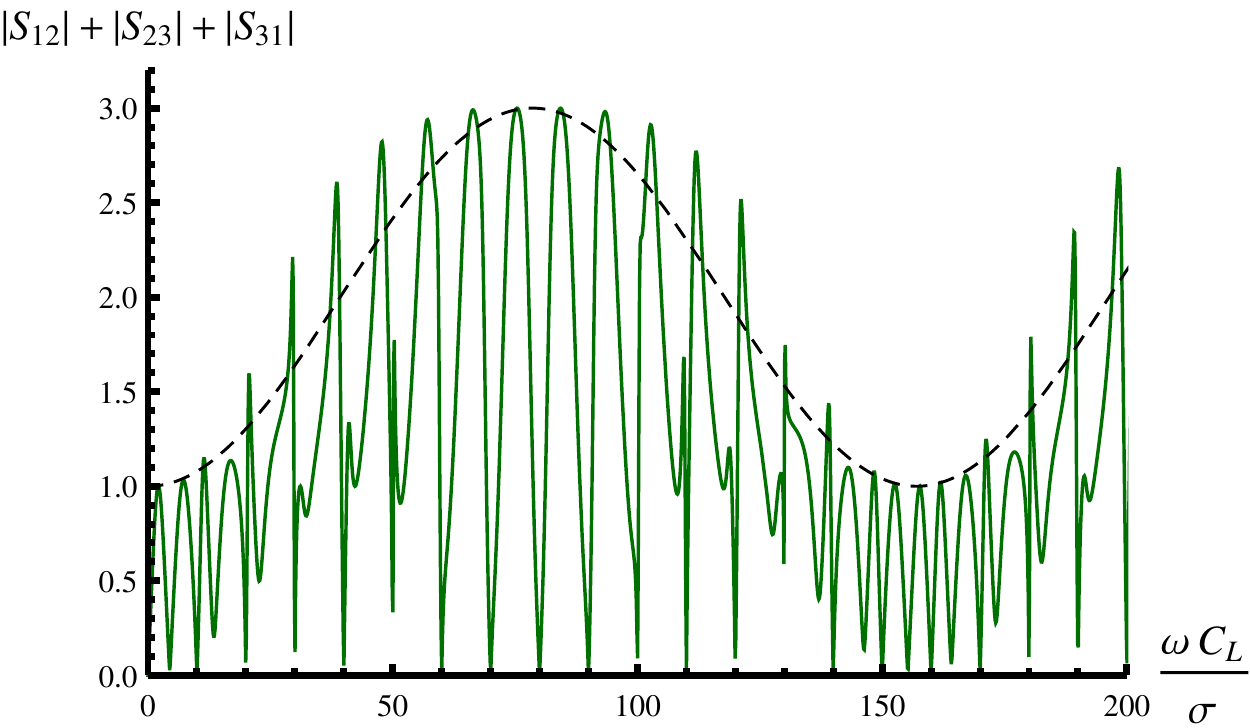}\\
\caption{Characterization of the scattering matrix of the first Carlin construction Fig. \ref{fig:ccarlin1} with rounded capacitances.  Rounding is as in Eq. (\ref{taper}) with $\lambda=L/12$.  Upper panel: $\vert (S_{3T,\lambda})_{13}+(S_{3T,\lambda})_{21}+(S_{3T,\lambda})_{32}\vert$. Due to unitarity the maximal value this sum can attain is 3, and at such a point perfect counterclockwise ($1\rightarrow 2\rightarrow 3\rightarrow 1$) circulation is achieved.  We see that when the slow modulation due to the rounding begins to occur, perfect gyration is lost.  Bottom panel: $\vert (S_{3T,\lambda})_{31}+(S_{3T,\lambda})_{12}+(S_{3T,\lambda})_{23}\vert$.  We see that the modulation causes this quantity to periodically attain the value 3, meaning that perfect {\em clockwise} circulation ($1\rightarrow 3\rightarrow 2\rightarrow 1$) is attained.}
\label{fig:3tS}
\end{figure}

\begin{subequations}
\begin{eqnarray}
Y_{3T,\lambda}=&&
\left(
\begin{array}{ccc}
 i a_\lambda & b_\lambda & -b_\lambda^* \\
 -b_\lambda^* & i a_\lambda & b_\lambda \\
 b_\lambda & -b_\lambda^* & i a_\lambda \\
\end{array}
\right)\,,\\
a_{\lambda}=&&2\sigma\frac{\sin \left(\frac{c \omega  (\lambda +L)}{\sigma }\right)-\sin \left(\frac{c \lambda  \omega }{\sigma }\right)}{1+2 \cos \left(\frac{c \omega  (2 \lambda +L)}{\sigma }\right)}\,,\\
b_{\lambda}=&&\sigma\frac{\exp(\frac{-i c \lambda  \omega }{\sigma }) \left(-1+\exp(\frac{-i c L \omega }{\sigma })\right)}{1+2 \cos \left(\frac{c \omega  (2 \lambda +L)}{\sigma }\right)}\,.
\end{eqnarray}
\end{subequations}
This response again has the same slow modulation in frequency as in the four-terminal case.  In Fig.~\ref{fig:3tS} we characterise the quality of the resulting impedance-matched circulator by computing the quantity  $\vert (S_{3T,\lambda})_{13}+(S_{3T,\lambda})_{21}+(S_{3T,\lambda})_{32}\vert$ which, due to the unitarity of the $S$ matrix, can be equal to three only for the case of an ideal circulator (independent of references phases).  We see that at low frequency perfect functioning is obtained; the response is not as robust as in the two-port case, in the sense that when the modulation due to the rounding becomes important, circulation is degraded (to recur again at higher frequency).  Another interesting functionality emerges: as the anti-clockwise circulation degrades, clockwise circulation as measured by  $\vert (S_{3T,\lambda})_{31}+(S_{3T,\lambda})_{12}+(S_{3T,\lambda})_{23}\vert$ occurs, which becomes almost perfect in a range of frequencies.  Thus, we have a set of interesting alternatives for achieving circulation.  Compared with the Hogan circulator, the Carlin circulators are more flexible, but are more sensitive to capacitance rounding and do not work properly when there is an impedance mismatch.

\section{Inductively Coupled Hall effect gyrator}\label{indgy}

\begin{figure}[htp] 
\centering 
\includegraphics[scale=0.4]{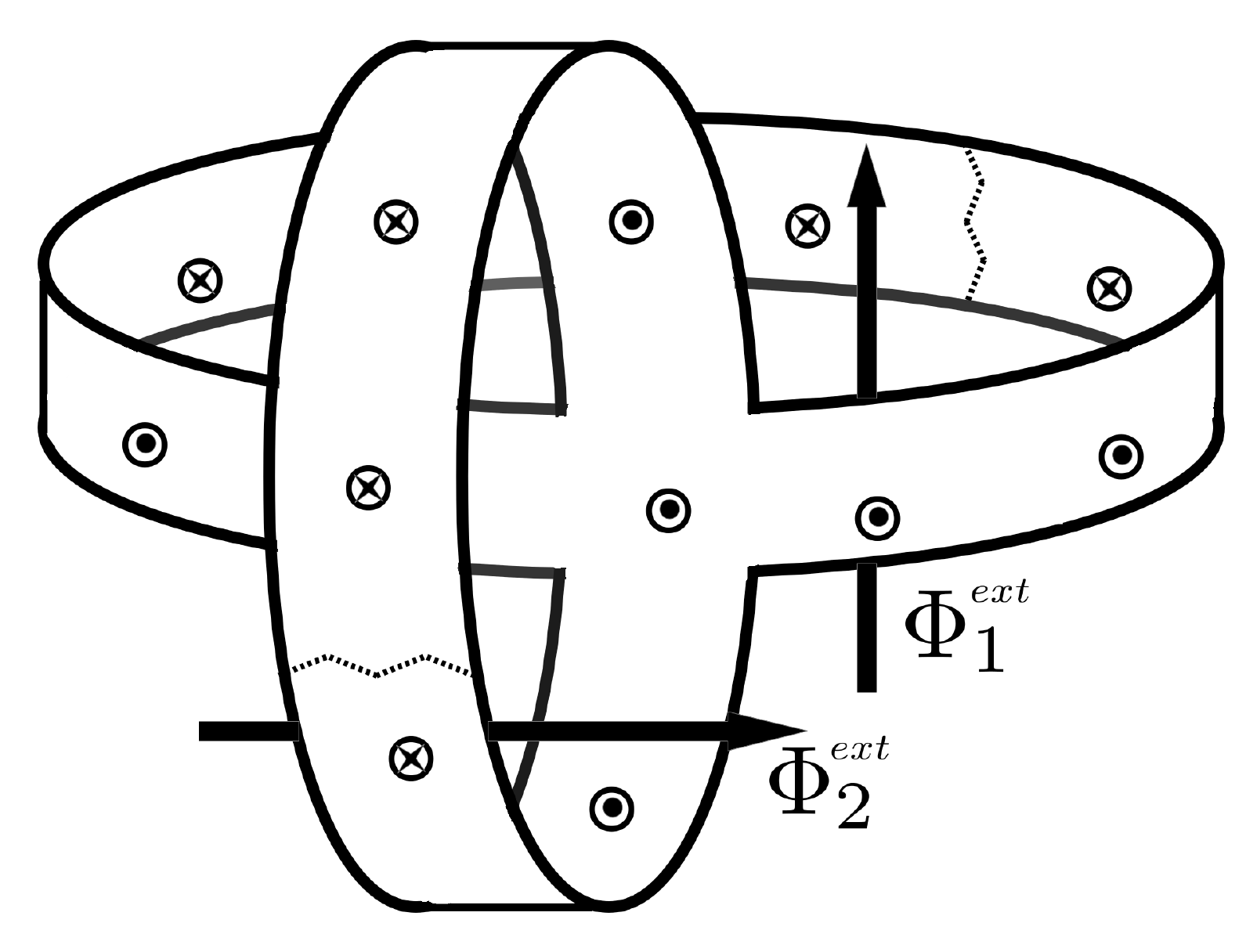}
\includegraphics[scale=0.4]{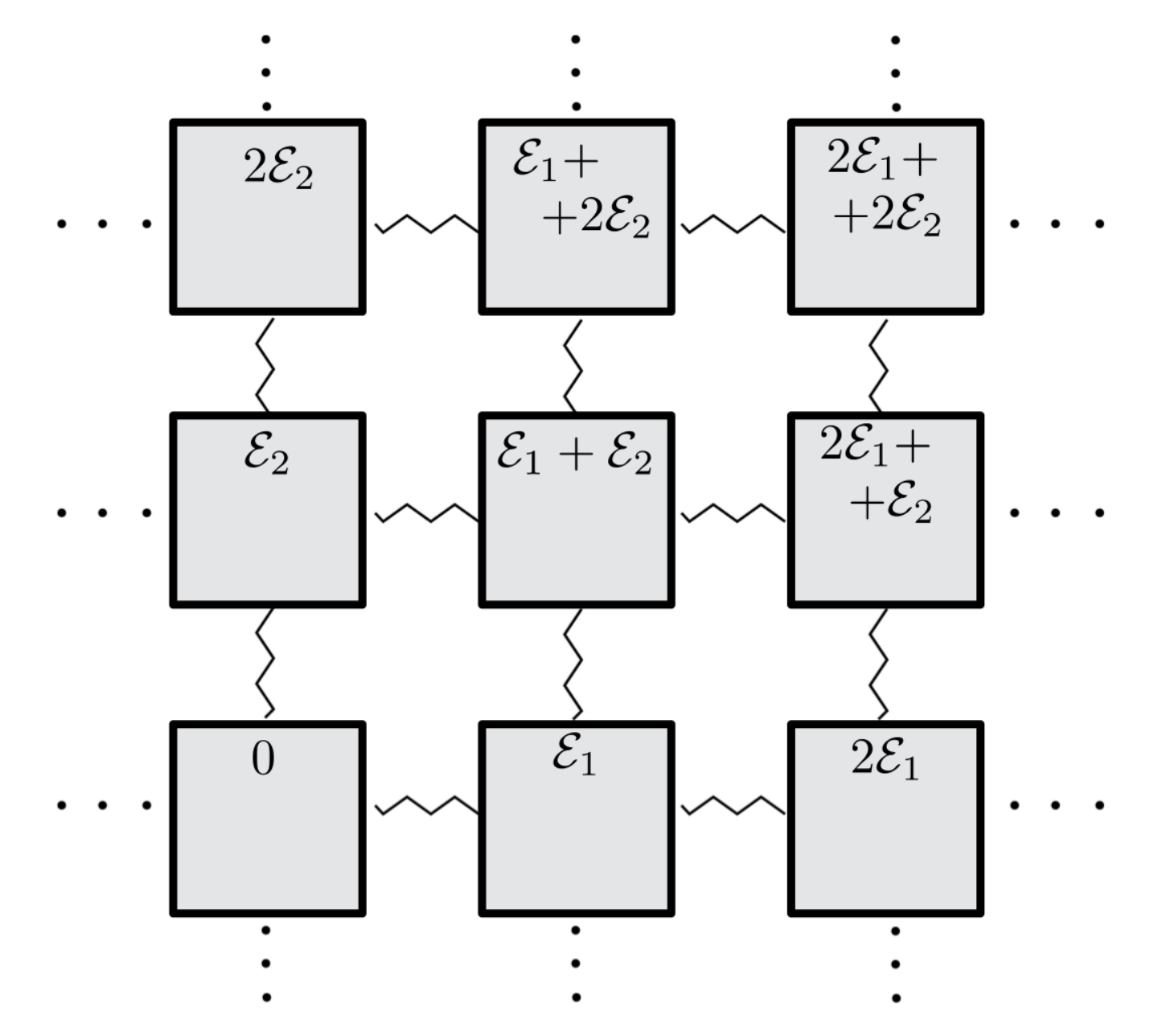}
\caption{Top: two-loop, nonplanar Hall conductor for realizing an inductively coupled gyrator.  The magnetic field texture necessary to produce the Hall effect in this conductor is shown; threading fluxes $\Phi_1^{ext}$ and $\Phi_2^{ext}$ can apply electromotive forces $\mathcal{E}_i=\dot\Phi_i^{ext}$ around the two loops. Bottom: periodic representation of conductor, in which the loops are unwrapped at the wavy lines in the top figure.  Solution to the field problem for Hall angle $\theta_H=\pi/2$ is given by setting the boundaries of this structure to the equipotentials indicated, which linearly increase from cell to cell as determined by the two e.m.f.'s $\mathcal{E}_{1,2}$.}
\label{fig:c-real}
\end{figure}

As pointed out in Tellegen's original work \cite{TellegenPRR}, both electric and magnetic effects can be considered for gyration.  We therefore briefly take up a dual approach to using the Hall effect for gyration, in which the lead coupling is magnetic rather than electric.  This approach leads to a very elegant view of the response of a Hall structure to magnetic induction, but we consider this approach less promising for application and will only give a sketch of the results.  

Inductive coupling requires loops of conductor, thus we consider the nonplanar Hall-material geometry shown in Fig.~\ref{fig:c-real}.  Topologically this surface is a torus with a hole cut into it; such a geometry was actually considered, for a Faraday material, by Tellegen in his later work~\cite{tellegenpatent2}.  While the topology we consider here has been standard in thought experiments for understanding the quantum Hall effect~\cite{Thouless}, and very analogous ``crossover'' Hall topologies have been noted for the achievement of interesting effects for quantum error correction~\cite{anyKitaev,Freedman52}, it must be understood that there is no material system in which there is a known technique for actually producing a material with a large Hall effect in such a topology.  It is for this reason that we do not anticipate that experiments can be performed to pursue this idea; but its principles are interesting to elucidate nonetheless.

Returning to Fig.~\ref{fig:c-real}, we suppose that one port provides input by the a.c. signal applied as a magnetic flux $\Phi_1^{ext}(t)$.  The time derivative of this flux produces an e.m.f. $\mathcal{E}_1$ around loop 1.  Since there is then a nonzero line integral of the electric field around this loop, the potential field $V(r)$ strictly speaking does not exist; but since $\nabla\cdot E=0$, one can locally define a potential that satisfies the Laplace equation; but it will be multivalued, increasing by $\mathcal{E}_1$ each time a path is taken around the loop.  This can be unwrapped into a periodic representation as shown in Fig. \ref{fig:c-real}.   In the limit of $\theta_H=\pi/2$ the boundary condition becomes trivial; we can use the equations given above with $c(s)=0$.  This says that the potential, in this periodic representation, is an equipotential on each of the periodic images of the loop edges as shown in Fig.~\ref{fig:c-real}.  Furthermore, the very simple relation between conductor current and boundary potentials, Eq. (\ref{curcalH}) means that the relation between the loop current and the e.m.f. in the other loop is perfect gyration, independent of frequency and dependent only on topology:
\begin{align}
I_2=\sigma{\mathcal E}_1, &&
I_1=-\sigma{\mathcal E}_2.
\label{loop}
\end{align}
The trouble with this approach, other than the extreme difficulty of producing non-planar conductors exhibiting a large Hall effect, is the need to couple externally to the variables of Eq. (\ref{loop}), which requires two transformer-like structures.  The weakness of magnetic coupling makes this problematic.  

We have found that the inductance $L$ of this coupling structure imposes a lower cutoff on the frequency at which gyration becomes effective.  This frequency scales like $\omega_{cutoff}\sim R_{gy}/L=1/\sigma L$.  The scale of inductance is set by $L\sim\mu\mu_0 d$, where $d$ is the physical scale of the device.  If the scale of $\sigma$ is the quantum scale $h/e^2$, then the cutoff, expressed as a wavelength, is given by the scale of $d\sim$~wavelength$/\alpha$, where $\alpha$ is the fine structure constant.  This suggests that the physical scale of the inductive device needs to be $\sim$137 times larger than wavelength of the a.c. radiation on which it operates.  The normal method of combatting this size penalty in transformer structures is to use high permeability materials (high $\mu$) and enhancing the inductance by multiple turns of conductor.  While this is a successful strategy for ordinary transformers, it is problematic here because the high permeability would need to be retained at high applied magnetic field (see following section), and, even worse, that the Hall conductor would need to be formed into some multi-turn corkscrew.  Given that even the one-turn structure of Fig. \ref{fig:c-real} is beyond any present capability, we would not judge these strategies for making an inductively coupled gyrator very promising.

It is worth noting that applying the same scaling argument to the capacitively coupled gyrator goes much more optimistically: the characteristic frequency goes like $\omega_{cutoff}\sim 1/RC=\sigma/C$, the scale of $C$ is $C\sim\epsilon\epsilon_0 d$ \footnote{This rough dimensional estimate applies both to both geometrical and quantum capacitance, see Sec. \ref{exgy}.}, so that if again we take $\sigma\sim h/e^2$, then we infer
\begin{equation}
d\sim\alpha\,\times\,{\mbox{\rm wavelength}},\label{reallysmall}
\end{equation}
that is, the natural scale of our capacitive device is 137 times {\em smaller} than the wavelength, that is, 137 times smaller than the natural scale of Hogan's Faraday-effect circulator.  This comparison is perhaps unfair, since the desired admittance scale of $1/50\Omega$ wipes out the factor of $\alpha$ from Eq. (\ref{reallysmall}); on the other hand, it is very easy to make capacitors whose capacitance far exceeds the dimensional estimate just used (viz., the parallel plate capacitor with area much larger than thickness), and we have seen that there are impedance-matching possibilities in the calculations given above so that, at least to achieve gyration over narrow bandwidths, matching $1/\sigma$ to $50\Omega$ need not be necessary. For engineering applications, the natural impedance-match condition $1/\sigma=50\Omega$ would, of course, be ideal.  Two routes are available for this: First, $\sigma$ can be some integer multiple $\nu$ of $e^2/h$; filling factor $\nu$ in the range of 10-20 is feasible. Second, a stack of Hall conductors can be put in parallel, further increasing the total conductance.  Of course, to keep the gyration frequencies in the desired range while increasing $\sigma$, the total capacitances would also have to be correspondingly increased (cf. Eq. (\ref{pgy})).   

\begin{figure}[htp] 
\centering 
\includegraphics[scale=0.3]{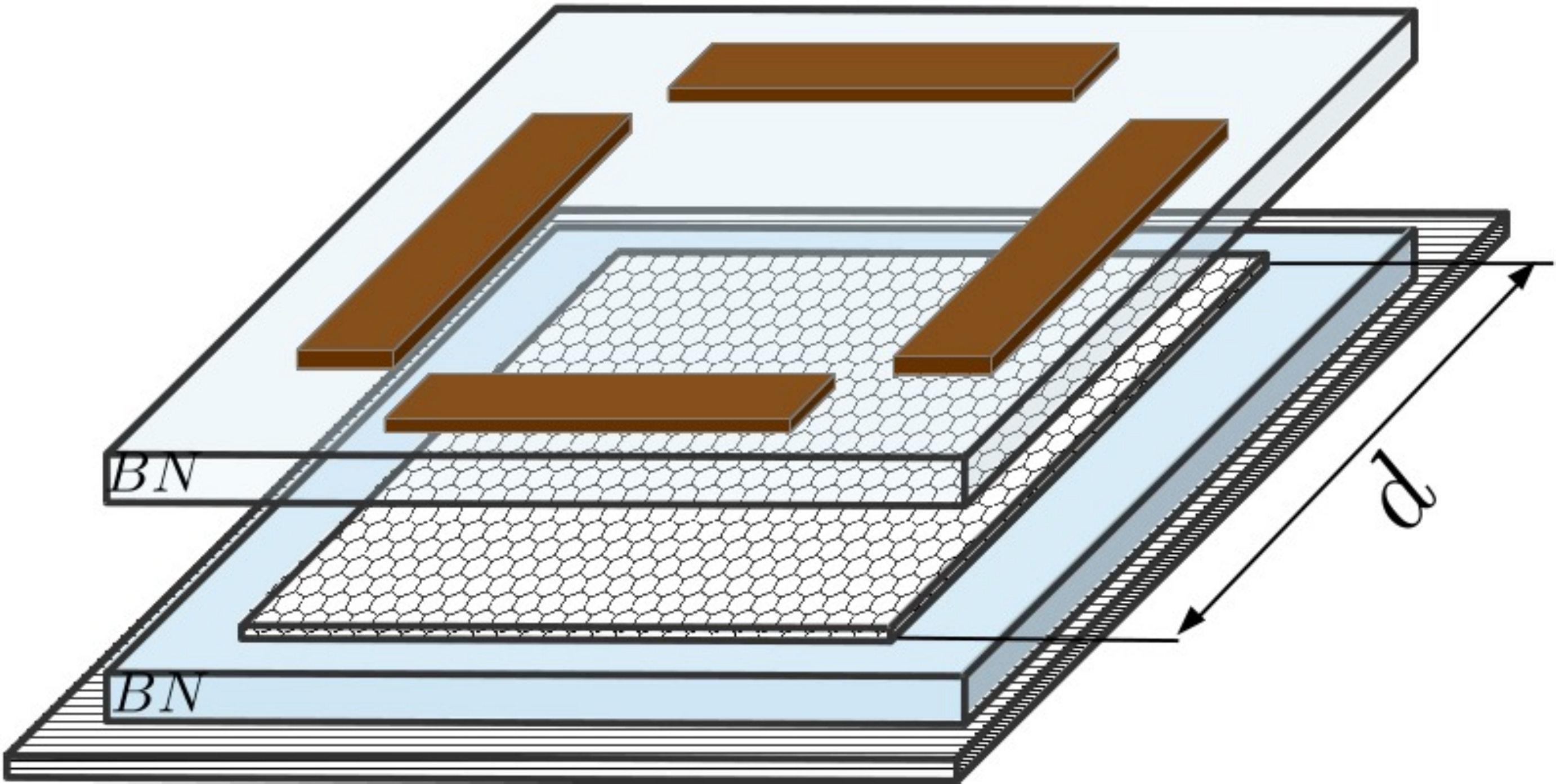}
\caption{An exploded view of a sandwich structure, based on the capabilities recently reported in~\cite{Wang01112013}.  A graphene flake is encapsulated between two layers of insulating boron nitride (BN).  Four edge electrodes grown above the structure as shown could serve as the four capacitive contacts of the two-port gyrator.
}
\label{fig:layers}
\end{figure}

\section{Experimental concepts for capacitive gyrator}\label{exgy}

Here we will explore the relation of our capacitive gyrator proposal to experimental observations in recent years involving magnetoplasmonic phenomena in Hall conductors, both in III-V heterostructures and in graphene.  Under conditions of the quantum Hall effect, $\theta_H=\pi/2$ (i.e., when the conditions Eq. (\ref{QHall}) apply), evidence for dispersionless edge magnetoplasmon propagation was already observed in the late 1980s \cite{Andrei,Gorkov}.  It was later proved \cite{PhysRevB.45.3894} that these magnetoplasmons propagate with low loss, and chirally (that is, in one direction only along the edge).  Further work established that the propagation velocity of these excitations follows a law like Eq. (\ref{vel}); however, the edge capacitance $c$ does not follow a simple classical picture.  In fact, it is quantitatively confirmed \cite{Gabelli28072006,PhysRevLett.98.166806} that the quantum capacitance picture, as analysed theoretically by B{\"u}ttiker and coworkers \cite{ButtikerQcap,Buttiker1993364,PhysRevLett.97.206804}, 
is necessary for explaining the observed dynamics.

The quantum effect involved in the quantum capacitance is the Pauli exclusion principle.  Unlike in an ideal classical metal, electric charge cannot be added or removed from the conductor without a change of the electrochemical potential.  This manifests itself as an extra effective capacitance, in parallel with the classical geometrical capacitance, given by the equation
\begin{equation}
C_q=e^2 {dN\over dE}.\label{cofq}
\end{equation}
Here $dN/dE$ is the density of levels around the Fermi energy.  In the ideal quantum Hall state this is quite small, so that $C_q$ is small and can easily dominate over the geometrical capacitance.   In this state there are no bulk states at the Fermi energy, so that only states at the edge of the conductor contribute.  Within the standard edge state picture, the edge state capacitance per unit length \cite{Gabelli28072006}, per edge state (corresponding to filling factor $\nu=1$), is \cite{PhysRevB.38.9375}
\begin{equation}
c_q={e^2\over h}{1\over v_{drift}}.\label{dri}
\end{equation}
Here $v_{drift}$, the velocity of the electron wave functions on the edge, has another simple classical meaning: it is the drift velocity of a ballistic direction subject to crossed magnetic and (confining) electric fields.  From Eq. (\ref{vel}) we see that if the edge capacitance is $c_q$, which will be true as long as the geometric capacitance is in excess of this modest value, then the magnetoplasmon velocity is essentially equal to the drift velocity.  This is a very special coincidence of the quantum chiral edge state situation, in general plasmon velocities and Schr{\"o}dinger wave velocities are determined by very different parameters.

There has been a very recent surge of interest in these investigations in the new graphene quantum Hall system.  The same chiral plasmon physics is also readily observed in this system \cite{Kumada}.  Precise magnetoplasmon parameters have recently been observed in graphene flakes \cite{PhysRevLett.110.016801}, with measured edge quantum capacitance per unit length found to be $c_q=100pF/m$, very consistent with theoretical estimates for graphene based on Eqs. (\ref{cofq},\ref{dri}).  The latest report of this work, has, in fact, clearly indicated the potential for graphene chiral magnetoplasmons for microwave circulators and other applications \cite{1310.6995P}.

The results of this paper indicate definite directions and design criteria that can put this realisation into practice.  To properly interface the plasmonic excitations, whose physics has now been well documented, with the in- and out-propagating guided electromagnetic waves of a real device, our results indicate that all contacts to the device should be capacitive, and not the combination of capacitive and ohmic contacts that are currently used in physical experiments.  Our results further indicate that the physical scale $d$ of the device (see Fig. \ref{fig:layers}), in order for there to be successful gyrator and circulator action in the GHz frequency range, should be in the millimetre range, given the measured values of $c_q$.  (There could be some advantage in going to III-V heterostructure Hall conductors; especially with soft edge confinement, the drift velocities can be smaller than in graphene, with a correspondingly larger $c_q$ and smaller length scale for the GHz device.)

According to our work, the optimal device would have most of its perimeter occupied by contact capacitors, to maximise $C_L$ and to minimise any stray capacitance of uncontacted edges to ground.  That is, all displacement currents should travel in and out of the conductor via the contact capacitors; as noted earlier, displacement currents to ground must be avoided.  Note that a gate (top or bottom), even one with a very large geometric capacitance, is not a concern here, since its quantum capacitance is virtually zero (because there is no bulk density of states, see Eq. (\ref{cofq})), so it will carry no ac displacement current.

We can mention one other scenario, in which the optimisation of the device structure would be quite different.  One can, with a very slight adjustment of parameters (e.g., magnetic field) work not in the fully developed quantum Hall regime, when $\theta_H$ is precisely 90 degrees, but rather in regime of non-maximal Hall effect, e.g., $\theta_H=85$ degrees.  This would make the device lossy, but, especially in the isolator application, some small degree of loss is not very detrimental to its operation.  In this regime, away from the quantum Hall ``plateaus'', bulk density of states is present, meaning that $dN/dE$, and $c_q$, is much larger.  Under these circumstances, an enhanced geometrical capacitance, achieved by making a top capacitor extending into the bulk of the conductor some distance from the edge (as suggested by Fig. \ref{fig:layers}), could lead to a much more miniaturised device.  Rough calculations suggest that GHz operation could then even be achieved for $d$ in the range of $d=10\mu m$.

At this length scale, a new encapsulation technique \cite{Wang01112013} indicated in the figure, which involves sandwiching an isolated flake of graphene between two extremely thin (c. 10nm) layers of insulating boron nitride (BN), has made available graphene samples with very small disorder (which could permit high Hall angle to be achieved for larger filling fraction $\nu$ and/or at higher temperatures).  It has been known for some time that the quantum Hall effect is rather robustly achievable in graphene, with $\sigma=h/e^2$ corresponding to one filled Landau level.  Larger $\sigma$, corresponding to filling multiple Landau levels, is also achievable, and would permit operation at smaller magnetic field. Magnetic fields on the Tesla scale will be required; one might speculate that micromagnet structures could permit a very compact encapsulated device with small fringing fields.

A small modification of the ohmic contacting technique pioneered in \cite{Wang01112013} should permit very well-controlled fabrication of the lead capacitors indicated in the figure.  A new difficulty would arise because, unlike in the fully developed quantum Hall situation, the bulk density of states would be nonzero and a gate capacitor would convey undesired displacement current in and out of the sample, depending on the details of the bulk charge transport mobility \cite{girvinrev}.  Thus, consideration would have to be given to making the bulk of the conductor floating, or controlled only by a very low $C_g$, remote gate capacitor.

\section{Conclusions and Outlook -- Quantum Effects}\label{conc}

While the use of Hall conduction for the achievement of gyration and circulation was declared impossible in 1954, the results of this paper indicates that this conclusion was premature; with current device capabilities, such a gyrator might actually be possible in the near future.  It is curious that the fundamentally different possibilities offered by reactive rather than galvanic coupling to the Hall conductor were not already examined a long time ago.  Capacitive coupling was always, of necessity, the method of contact for the two-dimensional electron gas (2DEG) formed by electrons floating on the surface of liquid helium.  But throughout the large literature on this subject~\cite{PhysRevB.50.11570,Peters1994674,0953-8984-5-22-010,PhysRevLett.67.2199,Lea1997}
 it seems that this coupling scheme \st{is} was always viewed only as a means to learn the basic response coefficients of this electronic system, rather than an interesting device feature in its own right. In metrological discussions~\cite{Ahlers09,Cage1998} careful accounting of capacitive effects has been made, but only in a setting where the basic coupling is ohmic.  Finally, there is other literature in which transport through semiconductor 2DEGs is achieved with capacitive coupling~\cite{Kim,Company2007}, but with the orientation that  the experimental data extractable from capacitive vs. ohmic contacts are equivalent, without any attention given to the difference that this might produce.

The present study is obviously incomplete, in that no quantum analysis has been provided for the functionalities that we have studied.  The classical Ohm-Hall approach has proven its worth in modelling the phenomenology of Hall-conduction devices from the 1950s~\cite{Mason1953} up to the present~\cite{PhysRevB.78.035416,Skachko13122010,PhysRevB.80.045408}.  While we can expect that some new quantum or mesoscopic phenomena would manifest themselves in the capacitively coupled devices that we have analysed here, perhaps at low temperature or in very clean systems, we can feel comforted that since the properties we have discussed here are fundamentally classical, they should be robust even in the face of considerable disorder, or at (moderately) high temperature. While the achievement of Hall angles very precisely equal to 90 degrees is very important in metrological applications, it is not so important here; a Hall angle of 85 degrees would still permit excellent gyrator, isulator or circulator action.

Quantum considerations are clearly very significant in setting limits on the validity of the results derived here.  The classical theory has no limit on the linearity of the response; we should expect departures from linearity at least when the potential drops in the device reach the Landau-level energy spacing.  Likewise, operating frequencies are certainly limited to below the inter-Landau level transition frequency.  In a classical theory any plasmon velocity is possible,with a straightforward geometric dependence on edge capacitance; the quantum description, as we have seen, intimately links the edge plasmon velocity to the drift velocity, itself fixed by the phase velocity of electron Schr{\"o}dinger waves. Finally, the classical theory has no lowest length scale of validity, while the quantum magnetic length is clearly a lower limit on the device dimensions that can reasonably be considered.

Fortunately, there is a strong basis for further work on the quantum aspects of this problem, as established in the theoretical work of B{\"u}ttiker and co-workers in the transmission theory of admittance and dynamic conductance~\cite{ThomasPretre,PhysRevB.53.2064}. Recent work 
 of Aita {\em et al.}~\cite{Aita2013} offers significant progress in defining the basic elements of a theory including electron correlation effects, going beyond the Hartree treatment of previous work.  Time will tell what tools will be needed to model important new aspects of this problem. 

While the Hall effect was declared unsuitable for the realisation of gyrators and circulators sixty years ago, we can hope that, after a long period of quiescence, the simple idea of reactive coupling to the Hall conductor will lead to a successful revival of this idea, with novel, miniaturised devices providing useful alternatives for constructing new  low-temperature quantum technologies. 

\section*{Acknowlegements}  
 
We are grateful to Fabian Hassler for indicating the Hall effect as a feasible way to produce gyration, and to Christoph Stampfer for many discussions of the potential experimental difficulties of our proposals.   This work has benefitted from valuable communications and discussions with Brian Anderson, Phil Anderson, Hendrik Bluhm, Efim Brener, Sourin Das, Andrew Houck,  Fran\c{c}ois Konschelle, Arne Ludwig, Steve Lyons, Bob Newcomb, Cedric Sodhi, Firat Solgun, Barbara Terhal, Bernat Terr\'es and Andreas Wieck.
We are grateful for support from the Alexander von Humboldt foundation. 

\bibliographystyle{apsrev4-1}
\bibliography{SemiconductorGyratorNotesREVTEX}

\end{document}